\documentclass[]{aastex631}
\usepackage[figuresright]{rotating}
\usepackage{subfigure}

\usepackage{graphicx}
\pdfoutput=1
\usepackage{hyperref}
\hypersetup{linkcolor=blue}

\definecolor{green2}{rgb}{0.04,0.7,0.13}
\newcommand*{\fnref}[1]{\textsuperscript{\ref{#1}}}

\begin{document}

\title{The Radiowave Hunt for Young Stellar Object Emission and Demographics (RADIOHEAD): \\A Radio Luminosity--Spectral Type Dependence in Taurus-Auriga YSOs}

\author[0009-0003-3708-0518]{Ramisa Akther Rahman}
\affiliation{Department of Astronomy, Yale University, 219 Prospect Street, New Haven, CT 06511, USA}
\affiliation{Center for Astrophysics, Harvard \& Smithsonian, 60 Garden Street, Cambridge, MA 02138-1516, USA}
\affiliation{William \& Mary, 200 Stadium Drive, Williamsburg, VA 23185, USA}

\author[0000-0002-4248-5443]{Joshua Bennett Lovell}
\affiliation{Center for Astrophysics, Harvard \& Smithsonian, 60 Garden Street, Cambridge, MA 02138-1516, USA}

\author[0000-0001-9605-780X]{Eric W. Koch}
\affiliation{Center for Astrophysics, Harvard \& Smithsonian, 60 Garden Street, Cambridge, MA 02138-1516, USA}

\author[0000-0003-1526-7587]{David J. Wilner}
\affiliation{Center for Astrophysics, Harvard \& Smithsonian, 60 Garden Street, Cambridge, MA 02138-1516, USA}

\author[0000-0003-2253-2270]{Sean M. Andrews}
\affiliation{Center for Astrophysics, Harvard \& Smithsonian, 60 Garden Street, Cambridge, MA 02138-1516, USA}

\author[0000-0002-5688-6790]{Kristina Monsch}
\affiliation{Center for Astrophysics, Harvard \& Smithsonian, 60 Garden Street, Cambridge, MA 02138-1516, USA}

\author{Dan Ha}
\affiliation{Portland State University, 1825 SW Broadway, Portland, OR 97201, USA}



\begin{abstract}
We measure the radio continuum fluxes at the locations of all {\it Gaia}--confirmed members of Taurus--Auriga using {\it Karl G. Jansky Very Large Array Sky Survey} data (VLASS; 2--4\,GHz, $\sigma_{\rm{VLASS}}{\sim}110{-}140\,\mu$Jy, $2.5''$ resolution) spanning 3 VLASS epochs (2019, 2021, and 2023).
We present 35 detections coincident with young Taurus--Auriga stars (29 in individual VLASS images, 6 via stacking).
We find a strong dependence on spectral type, wherein the fractional detection rate of radio emission coincident with early-type young stellar objects (YSOs) is systematically higher than late-type YSOs, ranging from $25\%$ $\pm$ 13\% for B--F YSOs, 21\% $\pm$ 11\% for G YSOs, 18.4\% $\pm$ 6.3\% for K0--K4 YSOs, 15.5\% $\pm$ 5.4\% for K5--K9 YSOs, 7.0\% $\pm$ 2.7\% for M0--M2 YSOs, 2.3\% $\pm$ 0.9\% for M3--M6 YSOs, and 1.9\% $\pm$ 1.9\% for YSOs with SpTs later than M7.
We present cumulative density distributions of radio luminosity densities that demonstrate a significant luminosity enhancement for early- versus late-type YSOs. We find 25\% of the detected sources to be significantly variable.
We discuss possible interpretations of this dependence, which may reflect stellar magnetic activity, binary interactions, or stellar flaring.
We find that mid-infrared YSO class is a strong indicator of radio detectability consistent with higher frequency Taurus-Auriga VLA surveys, with class~III stars detected at a rate of 8.8\% $\pm$ 1.6\%, class~IIs at 2.0\% $\pm$ 1.2\%, and combined class~0s, Is and Fs at 8.0\% $\pm$ 5.4\%.
\end{abstract}



\section{Introduction} \label{sec:intro}
High-resolution observations of young stellar objects (YSOs) at radio wavelengths provide a unique view of the star and planet formation process.
These observations provide access to emission from circumstellar matter and disk winds, to trapped ionised particle emission around giant planets, right down to processes occurring on the surfaces of stars associated with coronal activity, accretion and mass outflows \citep[via but not limited to, for example, thermal continuum, free-free (or \textit{Bremsstrahlung}), and gyro-/synchrotron emission, see e.g.,][]{Wilner00, Gudel02, Rodmann06, Pascucci12, Beltran16, Rodriguez2017, Anglada18}.

Indeed, past surveys with the \textit{Karl G. Jansky} Very Large Array (VLA) have revolutionised our understanding of nearby star formation.
The first generation of VLA surveys offered fundamental insights on the radio brightness distributions of YSOs in terms of their accretion properties \citep[i.e., as these evolve from Classical T Tauri Stars, CTTSs, to Weak-Line T Tauri Stars, WTTSs, see e.g.,][]{Bieging1984, Kutner86, NealFeig1990, Philips91, White92}.
More recently, the Gould Belt surveys \citep[of Ophiuchus, Serpens, Orion, Taurus, Serpens, and Perseus;][respectively]{Dzib13, Kounkel14, OrtizLeon15, Dzib2015, Pech16} presented detections of hundreds of YSOs in the nearby Galaxy.
These works investigated the population-level trends at radio wavelengths for YSOs across all evolutionary stages, from the initial Class 0/I/F stage, through to Class III \citep[see the classifications presented in][]{Williams2011}. While late-stage YSOs are typically much fainter in infrared and millimeter wavelengths, the VLA Gould Belt Surveys revealed that this is not necessarily the case at radio wavelengths, where Class III YSOs can be as luminous as (or more luminous than) their earlier-stage counterparts.


The {\it Gaia} satellite \citep{Gaiadr1_2016, Gaiadr2_2018, Gaiaedr3_2021, Gaiadr3_2023} is now providing breakthroughs in our understanding of the membership of nearby star-forming regions.
With each public {\it Gaia} data release have come new, kinematic membership directories of young stellar objects, which have confirmed/rejected previous candidates, but overall increased the total sample of YSOs within several hundred parsecs by many 100s--1000s \citep[see, for example,][]{Galli2019, Galli2020, Luhman2020, Luhman22a, Luhman22b, Esplin22}. 
This is true even for very nearby regions of the Galaxy, for example, the low-mass star-forming region Taurus-Auriga at a distance spanning just 110--200\,pc, which recent \textit{Gaia} analyses have demonstrated hosts many hundreds more young stellar members than previously understood \citetext{e.g., by comparing \citealt{Luhman2010,Kraus2017} with e.g., \citealt{Esplin19, Krolikowski2021, Luhman23}}. 
In conjunction, with the public release of all-sky high-resolution radio mapping in the form of the VLA Sky Survey \citep[VLASS;][at 2--4\,GHz (S-band), with a median RMS sensitivity of $120\,\mu$Jy, covering the whole sky from $-40^{\circ}$ northwards, at $2.5''$ angular resolution in the VLA's B configuration]{Lacy2020}, the potential to systematically examine the radio luminosity distributions for complete star-forming regions has opened up.
Consequently, VLASS opens up the opportunity to sample new nearby, young stellar populations, as revealed by \textit{Gaia}.

In this paper we focus our attention on Taurus-Auriga, given its relative closeness and thus high potential for detecting radio emission from YSOs.
Taurus is amongst the best-studied star forming regions at mid-infrared and millimeter/sub-millimeter wavelengths \citep[for example,][]{Furlan2009, Andrews2013, Esplin2014, Long2018}.
Taurus-Auriga has also been observed at radio wavelengths with the VLA, for example \citet{Dzib2015}, where detections of 59 sources co-located with known YSOs are presented, at a sensitivity approximately an order of magnitude deeper than VLASS. 
Although the VLA has observed many Taurus YSOs already, many of the new 100s of Taurus members discovered by {\it Gaia} were outside of the VLA field of view (FOV) and thus went unstudied. 
Given that the increase in Taurus YSO members is dominated by new class~III members, we therefore seek to understand with these updated populations if any present as radio-bright, and if there are any correlations with stellar properties.

In this paper, we present our results and analysis from cross-matching the positions of all sources in the {\it Gaia} membership list of \cite{Krolikowski2021} with VLASS `Quick Look' images \citep{Lacy2020} for VLASS epochs 1.1/1.2 (1), 2.1/2.2 (2), and where present 3.1/3.2 (3). 
We selected the catalog of \cite{Krolikowski2021} (adopted without modification), which was the latest available Taurus-Auriga source list when this study was commenced.
We provide a description of our methodology and describe our data sets in §\ref{sec:obsmeth}, and discuss our results in §\ref{sec:results}.
We present detections of 35 {\it Gaia}-associated YSO sources and (for those assessed as bona fide Taurus Auriga YSOs) how these relate to their stellar demographics (e.g., stellar spectral type, YSO class, age, sub-region, binarity) and their typical variability.
We conclude and summarise our work in §\ref{sec:conclusions}, providing insight into next steps to further the analysis presented here.

\section{Observations and Methods} \label{sec:obsmeth}

\begin{figure}
    \centering
    \begin{minipage}{1.0\textwidth}
    \centering
    \includegraphics[width=1.0\textwidth]{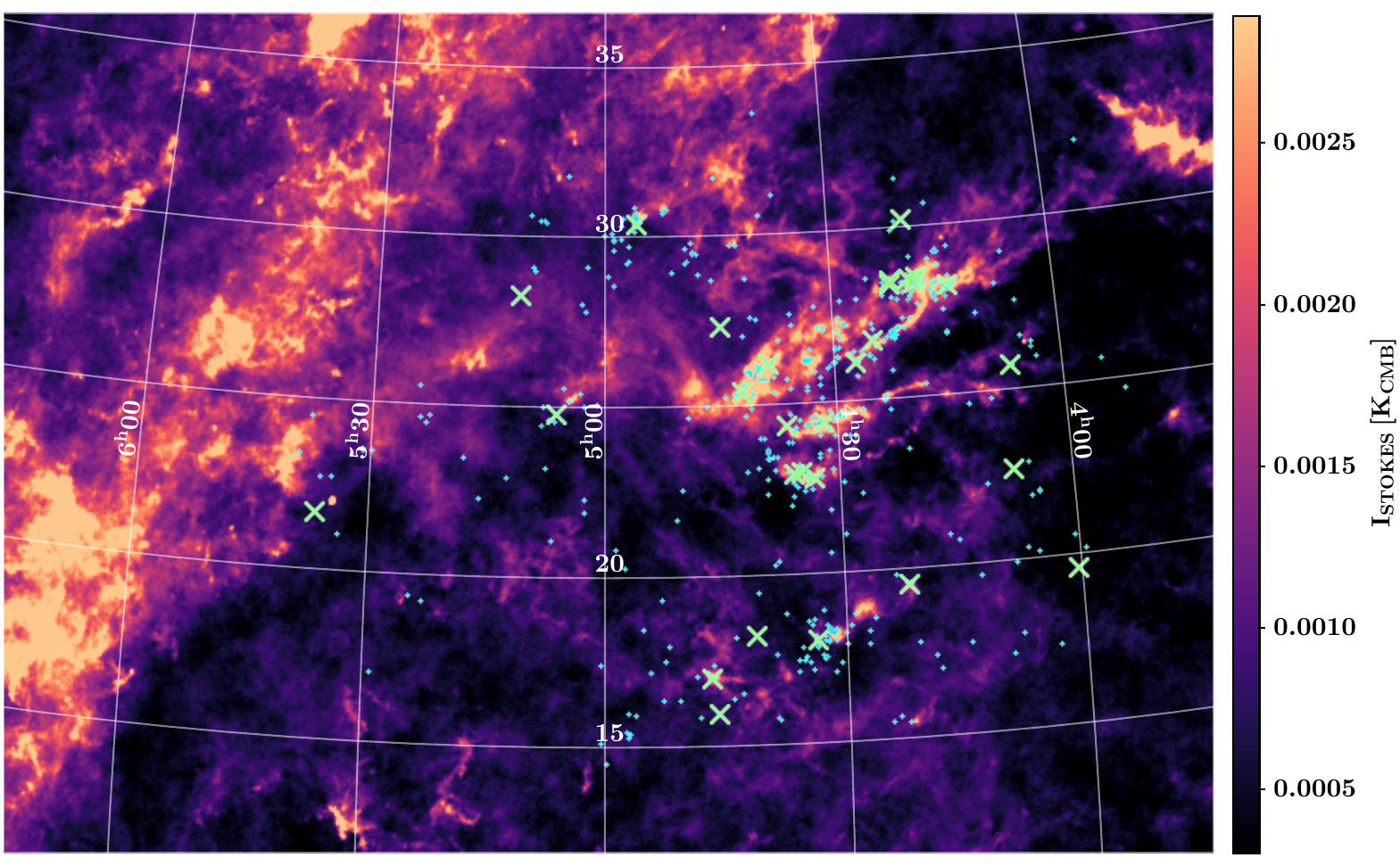}
        \caption{A \textit{Planck} (PR1) thermal dust map \citep[sourced from the IRSA database;][]{PlanckPR1} spanning the Taurus-Auriga complex. We highlight the locations of all 587 YSOs identified by \citet{Krolikowski2021} with cyan plus symbols, and those detected in VLASS data as larger green crosses (discussed in detail in \S\ref{sec:results}).}
        \label{fig:surveymap}
    \end{minipage}
 \end{figure}

\subsection{VLASS data extraction and inspection}
We extract all VLASS `QuickLook' (QL) images coincident with the 587 Taurus-Auriga YSOs as presented by \citet{Krolikowski2021}.
We use the Canadian Astronomy Data Centre's (CADC) service with the multi--epoch cut--out code, SODA\footnote{\url{https://gitlab.nrao.edu/mlacy/vlass-vo/-/blob/main/SODA\_multi\_pos\_multi\_ep.ipynb}} to create cut-outs of multi-epoch VLASS data.
All VLASS data were collated on 2023 June 01.
We plot all 587 sources on Fig.~\ref{fig:surveymap} on top of a \textit{Planck} (PR1) thermal dust map \citep[sourced from the IRSA database,][]{PlanckPR1} of the Taurus star-forming region.
QL images are calibrated using an automated \texttt{CASA} pipeline, which uses a flagging process that removes data corrupted due to factors such as radio-frequency interference (RFI) and VLA instrumental errors \citep{Lacy2020}.
VLASS QL images are science-ready and quality-assured data products provided by the NRAO. Caveats with their use as science products for analysis are outlined in VLASS memo 13\footnote{\url{https://library.nrao.edu/public/memos/vla/vlass/VLASS\_013.pdf}\label{vlass_memo}}, which highlights that the flux densities of sources brighter than 1 Jy should not be used, and that due to baseline coverage, flux recovery for structures extended ${>} 58''$ are unreliable. In our survey, all detected sources have flux densities ${<}$ 5.5 mJy, and are all point sources, thus are not impacted by these two major issues.
For VLASS epochs `1.2' onwards (i.e., for epochs 1.2, 2.1, 2.2, and 3.1), the total flux density is accurate to within 3\% of the Perley-Butler absolute flux density scale \citep[][{see also VLASS Memo 13}\fnref{vlass_memo}]{PB2017}.
This memo states that epoch 1.1 QL images have total flux densities low by 10\%, however all of the detections in our survey are made in epochs 1.2, 2.1 and 2.2, thus come with reliable photometry, and none of the extracted fluxes require any re-scaling.
Typical VLASS systematic positional errors are $\approx0.2''$ (for sources with declinations north of $-30^\circ$), and VLASS random positional errors are $\approx (1.5/\rm{SNR})''$, which in combination for any reasonably detected VLASS source, e.g., one with a SNR exceeding 4, is less than half of the $1''$ pixel size used in the VLASS imaging pipeline). 
Finally we note that a number of recent studies have taken advantage of these data already to study nearby stars as done in this study \citep[e.g., see][]{Davis2024, De2024}.

We generate images as $60''\times60''$ fits files, guaranteeing that each source is present within each file, whilst providing sufficient image area to assess local noise conditions. We additionally inspected the data products and excised images that did not meet certain criteria.\footnote{Upon manual inspection, we find that in 177 images, data for our sources is either located at the edge of or external to image files, all of which failed to meet our cutout threshold image size ($60''\times60''$, centered on source) thus we automatically excised these.
22 sources were returned with multiple images within the same epoch.
However, upon discussion with the VLA/NRAO Helpdesk (ticket ID: 36508) we found that images located within the same VLASS `tiles' were imaged with the same data, resulting in our choice to further excise the second SODA-returned QL image products where these had the same tile ID.
This affected all but 17/22 sources which we excised (i.e., 34 further images) resulting in $177+34=211$ total images rejected from our initial sample of 1408.
For 5/22 of these sources, independent data were present in all five different VLASS epochs, which we utilise in this work.
This overall resulted in 1197 separate images.}
The totals in each sub-epoch are as follows: epoch 1.1 (15), epoch 1.2 (575), epoch 2.1 (15), epoch 2.2 (577), and epoch 3.1 (15), and in the full epoch 1 (590), epoch 2 (592), and epoch 3 (15). 
There is a difference of 2 between the 1.2 and 2.2 epochs as there was one source that only had a usable epoch 2.2 image (no 1.2 image), and 1 source out of the 5 independent sources that had a single epoch 1.2 image and 2 epoch 2.2 images. 
In comparison to the \citet{Krolikowski2021} list of YSOs, 99\% (586/587) sources have images included in epoch 1 (epoch 1.1 and epoch 1.2 combined), 587/587 sources in epoch 2 (epoch 2.1 and epoch 2.2 combined) and 15/587 sources in epoch 3 (which only includes observations from epoch 3.1).

\begin{figure}
     \begin{minipage}{\textwidth}
        \centering
        \includegraphics[width=1.0\textwidth]{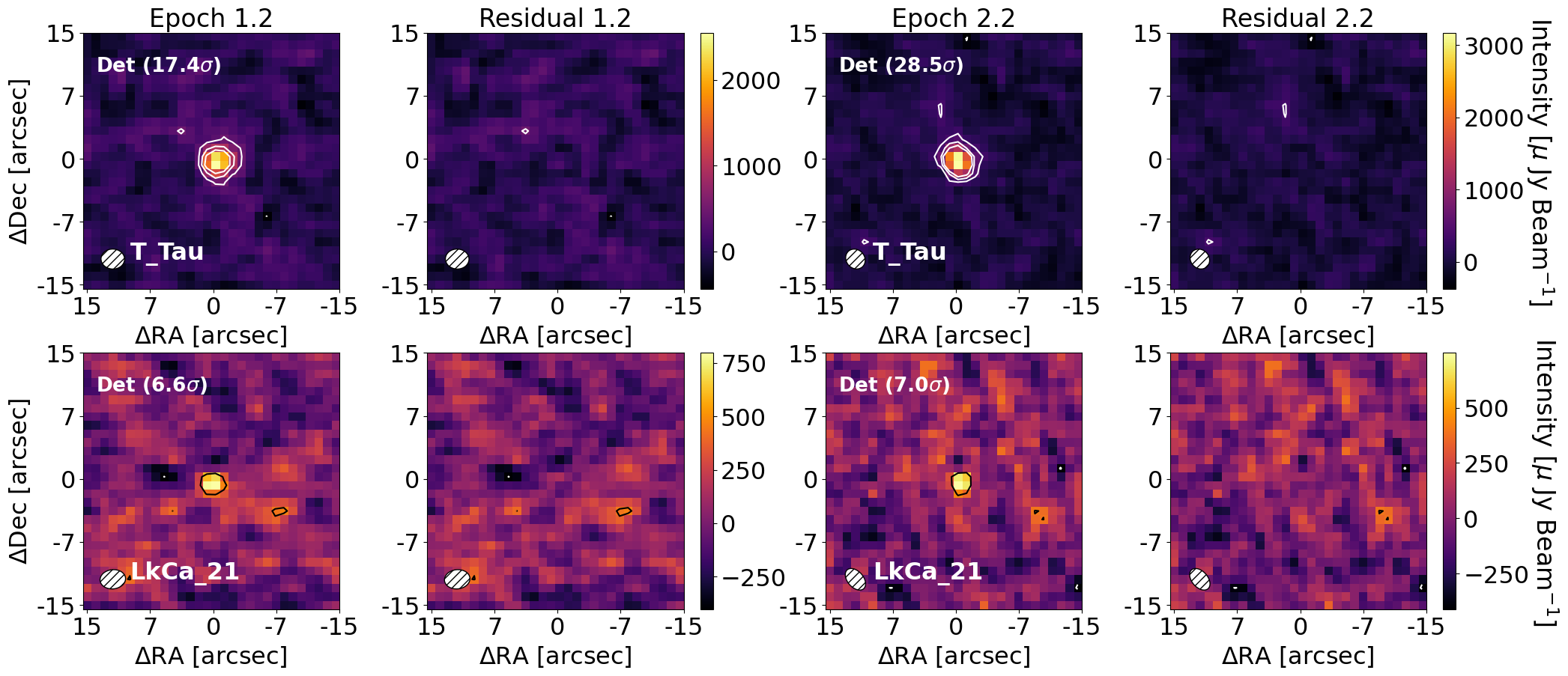}
        \caption{Example VLASS Quick Look epoch 1.2 and 2.2 images of detections, showing T~Tau and LkCa~21 as well as their residual maps. T~Tau has a detection significance in both epochs 1.2 and 2.2 of 17.4$\sigma$ and 28.5$\sigma$ respectively. LkCa~21 has a detection significance in epochs 1.2 and 2.2 of 6.6$\sigma$ and 7.0$\sigma$ respectively. VLA beams are in the lower-left. The $\sigma_{\rm{MAD}}$ in the epoch 1.2 and 2.2 images of T~Tau are 0.14\,mJy and 0.12\,mJy respectively. The $\sigma_{\rm{MAD}}$ in both the epoch 1.2 and 2.2 images of LkCa~21 are 0.13\,mJy. Contours are set at [-3,3,6,9]$\sigma$ levels. }
        \label{fig:VLASSexample}
    \end{minipage}
 \end{figure}

\subsection{VLASS fitting and detections}
We fit 2D Gaussian models to source locations with {\tt Astropy's Gaussian2D} class\footnote{\url{https://docs.astropy.org/en/stable/api/astropy.modeling.functional\_models.Gaussian2D.html}} \citep{2013A&A...558A..33A, 2018AJ....156..123A}.
These 2D Gaussian models are defined by an amplitude (brightness), a location (x, y), Gaussian widths (in major and minor axis directions), and a position angle.
We fix the Gaussian widths and position angles to match those of the image beams, and fit (positive) amplitudes located within one beam separation from each source's {\it Gaia}-determined position (equivalent to fitting unresolved point sources). We make no assumption regarding the possible angular scales of the sources. The data are well-fitted by these unresolved point source models, which correspond to beam-sized Gaussians in the image plane, and do not warrant fitting extended source models.
To estimate the errors of the fitted flux densities, we measure the median absolute deviation (MAD) per image. 
MAD statistics represent the average  deviation of fluxes from the median flux density (with most variation in the MAD caused by instrumental errors, weather variations and the presence of nearby bright emission sources), whereby if an image error is normally distributed, $\sigma_{\rm{MAD}}=1.4826{\times}$MAD.
The MAD statistic is less biased by outlier emission compared to the RMSE (Root Mean Square error). In RMSE, deviations from the mean flux are squared, which results in larger deviations being more highly weighted. Consequently, in images that suffer from bright source aliasing, the $\sigma_{\rm{MAD}}$ was found to provide a more reasonable estimate of the true noise (i.e., not skewed by emission unrelated to a source).
We set a ${>}4\sigma_{\rm{MAD}}$ detection threshold, whereby only converged solutions located within $2.5''$ of the {\it Gaia} source locations with an SNR exceeding this value are considered as detections (i.e., only sources within one beam from this location would be fitted).
We find in either epochs 1 or 2, that 29 sources present such detections (which we detail further in \S\ref{sec:results}).
We present two examples of VLASS detections coincident with YSO locations in Fig.~\ref{fig:VLASSexample}, showing detection images at both high-significance ({$17{-}29$}$\sigma$) and lower-significance ({$6{-}7$}$\sigma$), and their residual maps after subtracting their best-fit (unresolved) 2D Gaussian models, which in both epochs for both sources are consistent with noise.
Our best-fit model subtraction results in residuals that contain no excess emission exceeding $4\sigma_{\rm{MAD}}$.
For V773~Tau in epoch 1, we attribute residual emission to an imaging artefact that extends 10s of arcseconds in scale (as can be seen in Fig~\ref{fig:single_det_1}).
For sources with at least one source SNR ${>}3.5\,\sigma$, we stack both of their available epochs of data which yield images with uncertainties in the range 90--100 $\mu$Jy.
This results in 6 further detections with ${>}4\sigma_{\rm{MAD}}$. 

    \begin{sidewaystable}
    \centering
    \begin{tabular}{|l|l|cccccccccc|}
    \hline
    Source Name & {\it Gaia} EDR3 & Class & Age & SpT & GMM & Bin (sep.) & Flux$_{E1}$ & $\sigma_{\rm{MAD,\,E1}}$ & Flux$_{E2}$ & $\sigma_{\rm{MAD,\,E2}}$ & VAR SNR\\ 
    &&&[Myr]&&&&[$\mu$Jy]& [$\mu$Jy] (SNR)&[$\mu$Jy]& [$\mu$Jy] (SNR)&{($\delta$F)}\\\hline \hline
     HD~285281 & 50447448908539264 & III & 3.27 & \hyperlink{cite.Kraus2017}{F9} & D2 & -- & ${<}$570 & 140 & $660{\pm}120$ & 120 (5.4$\sigma$) & 0.5 \\ \hline
    V1195~Tau~B** & 162260226605718784 & III & 2.49 & \hyperlink{cite.Nguyen2012}{K6.5} & D4 & R2 ($1.0''$) & $860{\pm}140$ & 130 (6.4$\sigma$) & ${<}$480 & 120 & 2.1 \\ \hline
    J04073502$^\dag$$^A$ & 53515017631040384 & III & 6.22 & \hyperlink{cite.Herczeg2014}{M4.8} & D3 & -- & ${<}$530 & 130 & $690{\pm}130$ & 130 (5.2$\sigma$) & 0.8 \\ \hline
    V773~Tau & 163184366130809984 & II & 2.49 & \hyperlink{cite.Herczeg2014}{K4} & D4 & U & $5500{\pm}210$ & 130 (42.3$\sigma$) & $1800{\pm}140$ & 130 (14.3$\sigma$) & 14.7 {($3.1{\pm}0.3$)}\\ \hline
    V410~Tau & 164518589131083136 & III & 1.34 & \hyperlink{cite.Kenyon1995}{K3} & C2 & U & $1200{\pm}140$ & 130 (9.0$\sigma$) & $1500{\pm}140$ & 130 (11.8$\sigma$) & 1.6 \\ \hline
    V*~V1023~Tau & 164513538249595136 & III & 1.34 & \hyperlink{cite.Herczeg2014}{K8.5} & C2 & U & $1000{\pm}130$ & 130 (7.8$\sigma$) & $1400{\pm}130$ & 130 (11.2$\sigma$) & 2.1 \\ \hline
    V819~Tau & 164504467278644096 & III & 1.34 & \hyperlink{cite.Herczeg2014}{K8} & C2 & -- & ${<}$510 & 130 & $2200{\pm}150$ & 130 (16.8$\sigma$) & 8.7 {(${>}3.4$)} \\ \hline
    HD~281912 & 165327795329623936 & III & -- & \hyperlink{cite.Wichmann1996}{K1} & NM & -- & $670{\pm}140$ & 140 (4.9$\sigma$) & ${<}$470 & 120 & 1.1\\ \hline
    HD~283572 & 164536250037820160 & III & 1.34 & \hyperlink{cite.Herczeg2014}{G4} & C2 & -- & $2100{\pm}140$ & 130 (16.7$\sigma$) & $640{\pm}130$ & 130 (5.1$\sigma$) & 7.7 {($3.3{\pm}0.7$)} \\ \hline
    T~Tau & 48192969034959232 & 0IF & 1.73 & \hyperlink{cite.Herczeg2014}{K0} & C1 & U & $2500{\pm}160$ & 140 (17.4$\sigma$) & $3300{\pm}150$ & 120 (28.5$\sigma$) & 3.6 {($1.3{\pm}0.1$)} \\ \hline
    LkCa~21 & 164550882989640192 & III & 1.34 & \hyperlink{cite.Herczeg2014}{M2.5} & C2 & U & $880{\pm}140$ & 130 (6.6$\sigma$) & $900{\pm}130$ & 130 (7.0$\sigma$) & 0.1 \\ \hline
    V1201~Tau & 152104381299305728 & III & 3.09 & \hyperlink{cite.Wichmann1996}{K1} & C8 & -- & $810{\pm}130$ & 130 (6.1$\sigma$) & ${<}$470 & 120 & 1.9\\ \hline
    DG~Tau & 151262700852297728 & 0IF & 1.57 & \hyperlink{cite.Herczeg2014}{K7} & C6 & -- & ${<}$520 & 130 & $490{\pm}120$ & 120 (4.2$\sigma$) & 0.2 \\ \hline
    HK~Tau~A & 147847072275324416 & II & 1.57 & \hyperlink{cite.Herczeg2014}{M1.5} & C6 & R1 ($2.3''$) & $980{\pm}220$ & 220 (4.4$\sigma$) & ${<}$450 & 110 & 2.1 \\ \hline
    CFHT~Tau~12 & 145947077527182848 & III & 2.01 & \hyperlink{cite.Guieu2006}{M6.5} & C5 & -- & $570{\pm}140$ & 130 (4.2$\sigma$) & ${<}$470 & 120 & 0.6\\ \hline
    HD~28867~A & 3314244361868724224 & III & 3.27 & \hyperlink{cite.Walter2003}{B9} & D2 & -- & $1400{\pm}150$ & 150 (9.7$\sigma$) & $1900{\pm}140$ & 130 (14.5$\sigma$) & 2.1 \\ \hline
    XEST~08-003 & 145225596036660224 & III & 2.01 & \hyperlink{cite.Rebull2010}{M0} & C5 & -- & ${<}$510 & 130 & $1200{\pm}130$ & 120 (9.7$\sigma$) & 3.6 {(${>}1.5$)} \\ \hline
    HP~Tau~G2 & 145213192171159552 & III & 2.01 & \hyperlink{cite.Herczeg2014}{G2} & C5 & -- & $1400{\pm}130$ & 130 (10.9$\sigma$) & $1600{\pm} 130$ & 120 (13.8$\sigma$) & 1.3 \\ \hline
    J04363248$^\dag$$^B$ & 147614422487144960 & III & 2.49 & \hyperlink{cite.Luhman2017}{M8} & D4 & -- & ${<}$560 & 140 & $550{\pm}120$ & 120 (4.7$\sigma$) & 0.1 \\ \hline
    HV~Tau~A & 148450085683504896 & III & 2.59 & \hyperlink{cite.Herczeg2014}{M4.1} & C7 & U, R1 ($4.2''$) & $2700{\pm}180$ & 160 (17.3$\sigma$) & ${<}$480 & 120 & 10.5 {(${>}4.6$)}\\ \hline
    J04411296~A**$^{\dag}$$^{C}$ & 3409647203400743552 & III & 3.27 & \hyperlink{cite.Esplin19}{M4.75} & D2 & R1 ($2.2''$) & ${<}$560 & 140 & $560{\pm}120$ & 120 (4.6$\sigma$) & 0.02 \\ \hline
    LkHa~332~G2 & 148116246465275520 & III & 2.59 & \hyperlink{cite.Herczeg2014}{M0.6} & C7 & U & ${<}$560 & 140 & $490{\pm}120$ & 110 (4.3$\sigma$) & 0.4 \\ \hline
    HD~283782 & 154586318344905216 & III & 2.49 & \hyperlink{cite.Wichmann1996}{K1} & D4 & -- & $920{\pm}150$ & 150 (6.3$\sigma$) & ${<}$470 & 120 & 2.4 \\ \hline
    HD~30171 & 3309789385567829376 & III & 3.27 & \hyperlink{cite.Kraus2017}{G3} & D2 & -- & ${<}$520 & 130 & 1630 ${\pm}$ 170 & 160 (10.2$\sigma$) & 5.2 {(${>}2.2$)} \\ \hline
    DQ~Tau & 3406269842981827712 & II & 3.27 & \hyperlink{cite.Herczeg2014}{M0.6} & D2 & -- & ${<}$620 & 150 & $1000{\pm}140$ & 130 (7.6$\sigma$) & 1.9 \\ \hline
    HD~31305 & 156901168277114752 & III & 2.53 & \hyperlink{cite.Mooley2013}{A1} & C4 & -- & $990{\pm}140$ & 130 (7.4$\sigma$) & ${<}$480 & 120 & 2.8 \\ \hline
    CIDA~10 & 3419186939943738880 & III & 3.40 & \hyperlink{cite.Herczeg2014}{M4.2} & D1 & U & ${<}$550 & 140 & $600{\pm}110$ & 110 (5.3$\sigma$) & 0.3 \\ \hline
    LH98~175 & 3422476266419055232 & III & 2.34 & \hyperlink{cite.Li1998}{K0} & C3 & -- & ${<}$530 & 130 & $770{\pm}120$ & 120 (6.4$\sigma$) & 1.3 \\ \hline
    LH98~211 & 3403016495451584000 & III & 6.13 & \hyperlink{cite.Li1998}{K4} & C9 & -- & $750{\pm}170$ & 170 (4.3$\sigma$) & ${<}$540 & 140 & 0.9 \\ \hline
    \end{tabular}
    \caption{Statistics of YSOs detected with VLASS consisting of source name, YSO class, age, stellar spectral type, GMM, binarity (Bin), extracted flux per epoch, as well as the significance of source variability (`SNR var'). GMM refers to spatial subgroups of YSOs as identified by \citet{Krolikowski2021}, wherein each subgroup has a corresponding age (except in the case of sub-group NM). Bin indicates if a source has a resolved or unresolved binary companion, the number of companions, as well as the angular separation (if resolved). We denote unresolved binaries with ** which could indicate either (or both) A and B stars within these binary systems contribute to their radio emission. $^\dag$ indicate 2MASS ID sources (shortened for brevity) with the letters $^A$, $^B$, $^C$ referring to J04073502+2237394, J04363248+2421395, J04411296+1813194~A respectively. HD 30171 had one upper limit in epoch 3.1 with a flux of ${<}$$1400\mu$Jy (2s.f). References to all source SpTs are provided as embedded citations.}
    \label{tab:detections}
    \end{sidewaystable}

\begin{table}
    \centering
    \begin{tabular}{|l|l|ccccccc|}
    \hline
    Source Name & {\it Gaia} EDR3 & Class & Age & SpT & GMM & Bin (sep.) & Flux$_{\text{stk}}$ & MAD$_{\text{stk}}$ (SNR)  \\
        &&&[Myr]&&&&[$\mu$Jy]& [$\mu$Jy] (SNR)\\ \hline \hline
    GV Tau A & 149367383323435648 & -- & 2.49 & \hyperlink{cite.White2004}{K7} & D4 & R1 (0.5'') & 390 $\pm$ 91 & 90 (4.3$\sigma$)  \\ \hline
     V* V830 Tau & 147831571737487488 & III & 1.57 & \hyperlink{cite.Herczeg2014}{K7.5} & C6 & -- & 420 $\pm$ 100& 100 (4.1$\sigma$)  \\ \hline
    V* V1000 Tau & 148116177746232192 & 0IF & -- & \hyperlink{cite.Herczeg2014}{M2.5} & -- & U &380 $\pm$ 91 & 90 (4.2$\sigma$) \\ \hline
    J04554757$^\dag$$^A$& 156915878541979264 & III & 2.53 & \hyperlink{cite.Herczeg2014}{M5} & C4 & -- & 410 $\pm$ 88 & 86 (4.8$\sigma$) \\ \hline
    J05122759 A$^\dag$$^B$& 3415706130944329216 & III & 3.40 & \hyperlink{cite.Luhman2017}{M2.5} & D1 & R1 (0.8'') & 410 $\pm$ 94 & 92 (4.4$\sigma$) \\ \hline
    V* V1366 Tau & 3416236744087968768 & III & 6.13 & \hyperlink{cite.Kraus2017}{K7} & C9 & -- & 450 $\pm$ 98& 96 (4.7$\sigma$) \\ \hline
    \end{tabular}
    \caption{Statistics of YSOs detected with VLASS after stacking multiple epochs of data, consisting of source name, YSO class, age, stellar spectral type, GMM, binarity (Bin), and extracted stacked flux.  $^\dag$ indicate 2MASS ID sources (shortened for brevity) with the letters $^A$, $^B$, referring to J04554757+3028077 and J05122759+2253492 A respectively. References to all source SpTs are provided as embedded citations.}
    \label{tab:stacked_det}
\end{table}

\section{Results and discussion} \label{sec:results}
\subsection{Sample detection overview}
We observe significant (${>}4\sigma$) radio emission associated with 29 point sources coincident with the complete set of 587 source locations of \citet[][]{Krolikowski2021} in single images (i.e., a detection fraction of 4.9\%), and a further 6 sources in stacked images (i.e., a detection fraction of 6.0\% for the complete survey).
We present the detected source data, with derived fluxes and uncertainties (per epoch), and variability significance in Table~\ref{tab:detections}, ordered by right ascension, and for stacked sources in Table~\ref{tab:stacked_det}.
Table~\ref{tab:detections} only includes data from epochs 1 and 2 since in the incomplete epoch 3 data, only 1/29 of our sources (HD~30171) had an image which resulted in a non-detection. 

Where we present detection SNRs, these are measured only with respect to the noise (MAD) within their image data (and so have not been combined with calibration uncertainties).
Where we present absolute fluxes, we provide uncertainty estimates by combining image errors ($\sigma_{\rm{MAD}}$) in quadrature with an assumed 3\% calibration error. 
In epochs 1, 2 and 3 of a total of 586, 587 and 15 sources (present within the 590, 592 and 15 independent images, i.e., after removing multi-tiles, erroneous images, and dependently-imaged data, see \S\ref{sec:obsmeth} for details) we detect source-coincident emission in epoch 1 in 18/586 sources, in epoch 2 in 21/587 sources, and in epoch 3 in 0/15 sources.
In total we detect 39 point sources in these images, towards 29 distinct sources.
From the complete set of 1197 images (see \S\ref{sec:obsmeth}) we measure a sample mean uncertainty of 140\,$\mu$Jy and thus a mean detection threshold of 560\,$\mu$Jy (i.e., $4\times$ the sample mean uncertainty), and per-epoch mean detection thresholds of 600\,$\mu$Jy (epoch 1), 500\,$\mu$Jy (epoch 2), and 1000\,$\mu$Jy (epoch 3).
Although the total detection counts are broadly consistent, these sensitivity differences may reflect the (minor) increase in the epoch 2 detection counts.
Finally, the results indicate significant variability towards some sources (and consistent emission between epochs for others) which we discuss further in \S\ref{sec:variables}.

From the known source distances and determined fluxes, we calculate the luminosity densities of all sources ($L_R$, which we present in Table~\ref{tab:xraytable} as mean values). The total range of mean values of $L_R$ for detected sources is relatively narrow, from $6.4\times10^{15}$\,erg\,s$^{-1}$\,Hz$^{-1}$ (LH98~211) to $7.3\times10^{16}$\,erg\,s$^{-1}$\,Hz$^{-1}$ (T~Tau). In combination with the 6.0\% detection rate, these values overall suggest that the typical 3\,GHz radio luminosity densities of Taurus YSOs are low. Comparison with the more sensitive 4.5\,GHz and 7.5\,GHz surveys of Ophiuchus and Orion corroborates this assessment; the range we measure in Taurus is comparable with the brighter sources in Ophiuchus \citep{Dzib13}. These ranges are around an order of magnitude fainter than those of the younger, higher-mass star forming region, Orion \citep{Kounkel14}.

\subsection{Taurus YSOs: bright radio emission detected with VLASS}
Of the 587 sources in \citet{Krolikowski2021}, 59 are noted in their work as `non-members' \citep[{with GMM values denoted} `NM', see][which typically have {\it Gaia} astrometric solutions discrepant with the Taurus star-forming region, i.e., distances exceeding 225\,pc]{Krolikowski2021}, and a further 16 sources have no {\it Gaia} \textit{XYZ} data and thus cannot be confirmed as Taurus-Auriga star-forming region members. Of these (total) 75 sources, two are present in the 35 source detections in the VLASS data, HD~281912 and V*~V1000~Tau. In the case of HD~281912, this source is determined as having an astrometric solution with a distance of 370\,pc. In the case of V*~1000~Tau, this source has no available Gaia astrometric solution, and thus its membership as a Taurus-Auriga YSO cannot be confirmed. As such, we exclude HD~281912 and V*~V1000~Tau from further analysis in this work, alongside all such 75 sources in these same categories. This yields 512 YSOs that we refer to as the `full' sample for analysis purposes. A complete table including all statistics as those in Table~\ref{tab:detections} and Table~\ref{tab:stacked_det} for the full 512 sample is available in machine-readable form.

One caveat concerning the analysis of four unresolved binaries (2MASS~J04411296+1813194, V1195~Tau, GV~Tau~A/B, and 2MASS~J05122759+2253492) is that we cannot attribute the detection of radio emission specifically to either YSO A or B within the binary (i.e., either or both sources may contribute to the radio emission). 
In the case of 2MASS~J04411296+1813194 and V1195~Tau these are binaries with SpTs M4.75/M5.25 and K6.5/K6.5 respectively, which we list in Table~\ref{tab:detections} as M4.75 and K6.5 respectively. 
Whereas since the GV~Tau~A/B binary system hosts a K7 YSO (A) and an unknown SpT companion (B) and the 2MASS~J05122759+2253492 binary system hosts an M2.5 YSO (A) and an unknown SpT companion (B), we list in these cases the SpT of the primary within the binary/multiple.
Before proceeding further, we first discuss the variability in the sample, binarity of sources with detections, and consider the possibility that detections are due to coincident background sources.

\subsubsection{Contamination of emission from extra-galactic sources?}
We estimate the possible contamination rate of the detections by extragalactic sources using equation A11 of \citet{Anglada98}:
\begin{equation}
    N = 1.4\Big\{ 1 - \exp \left[-0.0066\Big( \frac{\theta_F}{\rm{arcmin}} \Big)^2 \Big( \frac{\nu}{5\,\rm{GHz}} \Big)^2 \right] \Big\}\Big( \frac{S_0}{\rm{mJy}} \Big)^{-0.75}\Big( \frac{\nu}{5\,\rm{GHz}} \Big)^{-2.52},
\end{equation}
for $N$ the number count of extragalactic sources, $\theta_F$ the total area within which such extragalactic sources could fall within, $\nu$ the observing frequency, and $S_0$ the sensitivity threshold for detecting such a source (this equation assumes an $\alpha=-0.7$ spectral slope for extra-galactic sources). We note that \cite{Dzib2015} likewise adopted this same expression to estimate their background contamination rates.
We calculate the expected number count ($N$) of contaminant extra-galactic sources as $N=1.7\times10^{-2}$ (i.e., for the mean VLASS frequency $\nu=3\,$GHz, detection sensitivity $\approx 600\,\mu$Jy, over the survey detection area of 1.1\,arcmin$^2$ from the 512 regions in which we fitted source emission).
This value of $N$ is low, which reflects the overall small area over which we fit fluxes, and the relatively low VLASS sensitivity.
Importantly, this measurement of $N$ implies that the detections are dominated by emission from YSOs and not extra-galactic background sources, i.e., there is a probability of around 1.7\% that just one source is an extra-galactic background object.

\subsubsection{Radio variability} \label{sec:variables}
We quantify the variability of the sample with two metrics, firstly, the SNR of the variability (in Table~\ref{tab:detections} `VAR SNR'), and secondly the flux factor by which the respective epoch flux values changed (in Table~\ref{tab:detections} `$\delta F$'). We quantify VAR SNR as $|F_{\rm{E1}}-F_{\rm{E2}}|/\sqrt{dF^2_{\rm{E1}} + dF^2_{\rm{E2}}}$. For the sources that show significant variability (at the ${>}3\sigma$ level), we quantify $\delta F = F_{\rm{i}}/F_{\rm{j}}$ (where $F_{\rm{i}}{>}F_{\rm{j}}$, i.e., the ratio of higher flux to lower flux). If a variable source was only detected in one epoch, then we quantify lower-limits as $\delta F = (F_{\rm{i}}-3\rm{e_{F_{i}}})/F_{\rm{j}}$, i.e., presenting these as $3\sigma$ lower-bounds. In total, we find 7/28 sources to be variable at the ${>}3\sigma$ level, and 5/28 sources to be strongly variable (at the ${>}5\sigma$ level). Discussing briefly in turn the sources we determine as variable, these are V773~Tau, V819~Tau, HD~283572, T~Tau, XEST~08-003, HV~Tau~A, and HD~30171. The radio variability of five of these sources has been reported elsewhere \citep[see e.g.,][for V773~Tau, T~Tau, V819~Tau, XEST~08-003, and HD~283572 respectively]{Massi2002,Johnston03,Dzib2015,Lovell2024}. To our knowledge, HV~Tau~A, and HD~30171 have not yet been reported as being variable sources at radio wavelengths.

Based on the mean sample uncertainties for all data in epochs 1 and 2 (i.e., 150\,$\mu$Jy in epoch 1, and 125\,$\mu$Jy in epoch 2), the VAR SNR metric implies that for a source to be deemed variable at the ${>}3\sigma$ level, typically this would require a change in flux at the level of ${\gtrsim}600\,\mu$Jy (where we assume image uncertainties to dominate over flux calibration uncertainties, true for all but the very brightest detections), which is corroborated by the values in Table~\ref{tab:detections}. Of the 7/28 variable sources, we find a wide range of $\delta F$ values, with lower-bounds ranging from 1.5 (XEST~08-003) to 4.6 (HV~Tau~A), and absolute values from $1.3{\pm}0.1$ (T~Tau) to $3.3{\pm}0.7$ (HD~283572). These ranges imply that these VLASS data are sensitive to modest variability from bright sources at the level of a few tens of percent (e.g. T~Tau), and strong variability from fainter sources at the level of a few hundreds of percent (e.g. HV~Tau~A).  Despite the shallow sensitivity of VLASS across just two-epochs from which we detect emission, the fraction of sources that are variable remains relatively high at 25\%, suggesting that the underlying sample variability is significant, and that we are not sensitive to low-levels of variability that may pervade the sample. Indeed, prior analyses of several targets from which we measure no significant variability found these to be radio variable \citep[for example, DG~Tau, DQ~Tau, and V410~Tau, see e.g.,][respectively]{Lynch13,Salter10,Golay23}. Overall, the finding that there is significant variability in the detected sample reflects an important possibility that the detections are dominated by stellar flaring (with emission arising from e.g., stellar rotation, or binary interactions).

\subsubsection{Binarity of detected sources} \label{sec:binariesDets}
Of the 33 detections, 13 are stated by \citet{Krolikowski2021} as being hosted in binaries/stellar multiples (39\% $\pm$ 8.5\%), in contrast to the 18.5\% $\pm$ 1.7\%  (95/512) rate of binaries/stellar multiples in the full sample.
Comparing these fractions shows discrepancy at the $2.4\sigma$ level, suggesting that the detected sample may be over-populated by binaries.
Indeed, it is known that binary interactions can cause periodic radio/millimeter emission (and variability), e.g., as shown for DQ~Tau and V773~Tau \citep[see e.g.,][]{Massi06,Salter10}, hence it may not be surprising to see a raised detection level of binaries/stellar multiples in the Taurus sample.
However, the plausible binary detection enhancement could simply reflect the binarity compilation criteria of \citet{Krolikowski2021} which excluded some stellar multiples.
We note for example that in the detected source list, DQ~Tau and HD~28867~A are known stellar multiples, yet have not been noted as such by \citet{Krolikowski2021}. 
This implies that the fractional rate of binaries in the sample and those noted as such in the detection table are both under-counting stellar multiples.
Our assertion is further supported by the fact that stellar multiplicity fractions are typically closer to 30--50\% for low-mass star forming regions \citep[see e.g.,][]{Duchene13}, significantly higher than the 18.6\% level stated above.
Since the binary statistics of \citet{Krolikowski2021} have been compiled in a consistent manner, we do not attempt to re-construct a complete table of binary systems in Taurus.
Therefore, whilst it remains plausible that binaries are over-populated in the sample of detections, we are unable to robustly demonstrate this with the available data. 
We return to discuss stellar multiplicity in the context of radio emission in our discussion of SpTs in \S~\ref{sec:SpTInterpret}.

\subsection{Radio luminosities as a function of stellar spectral type}
\label{sec:LR_spt}


\begin{figure}
\gridline{\fig{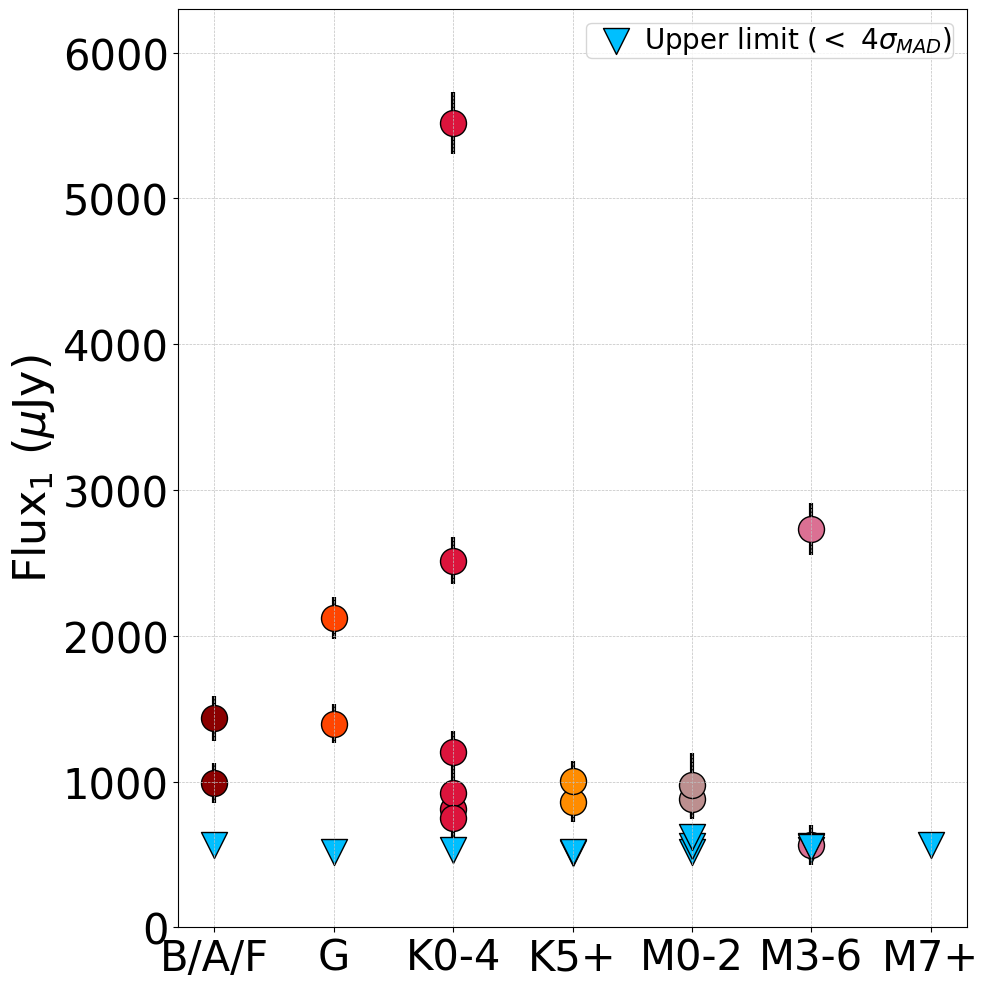}{0.48\textwidth}{(a)}
          \fig{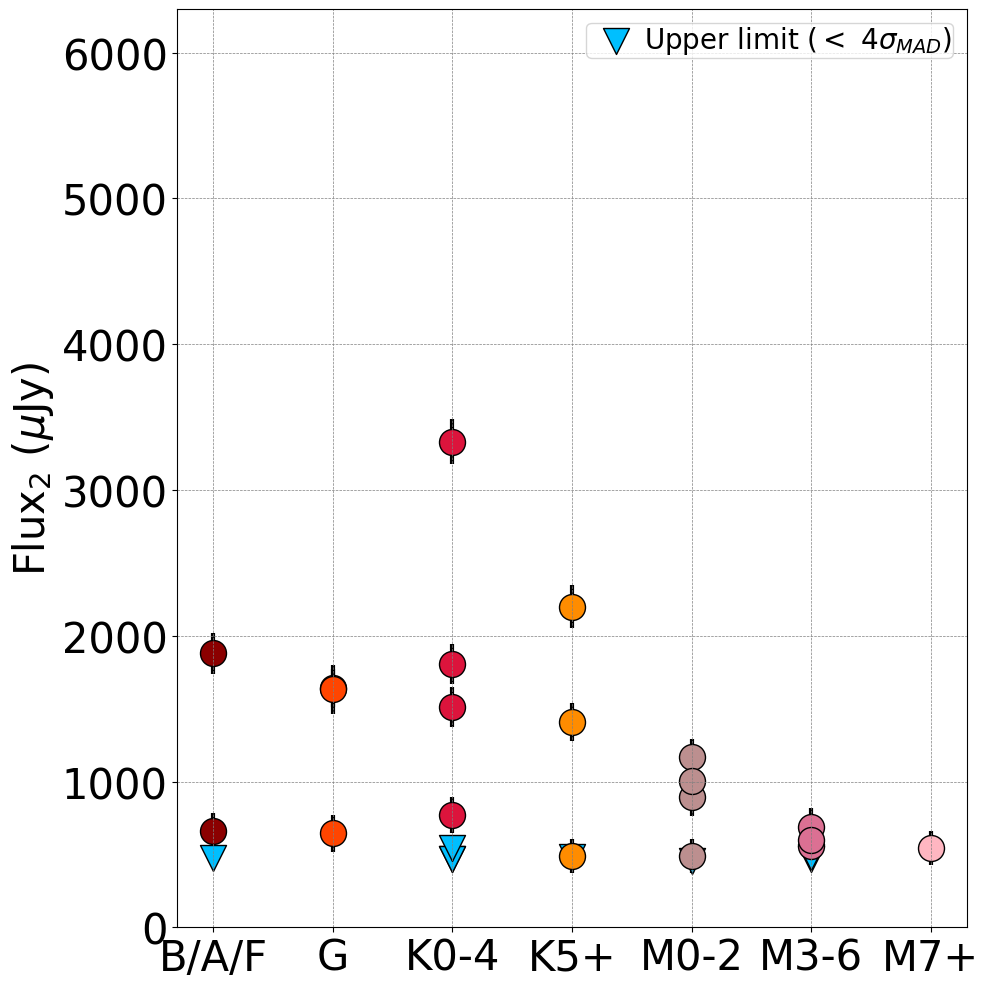}{0.48\textwidth}{(b)}}
\gridline{\fig{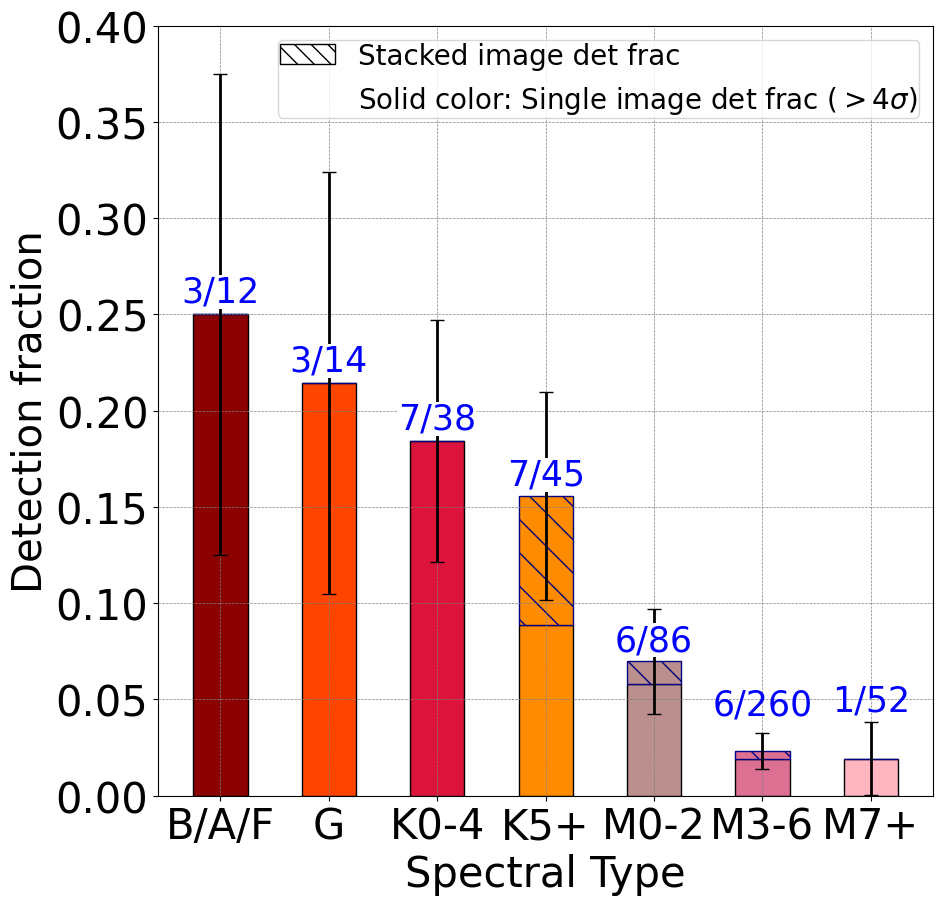}{0.5\textwidth}{(c)}
        \fig{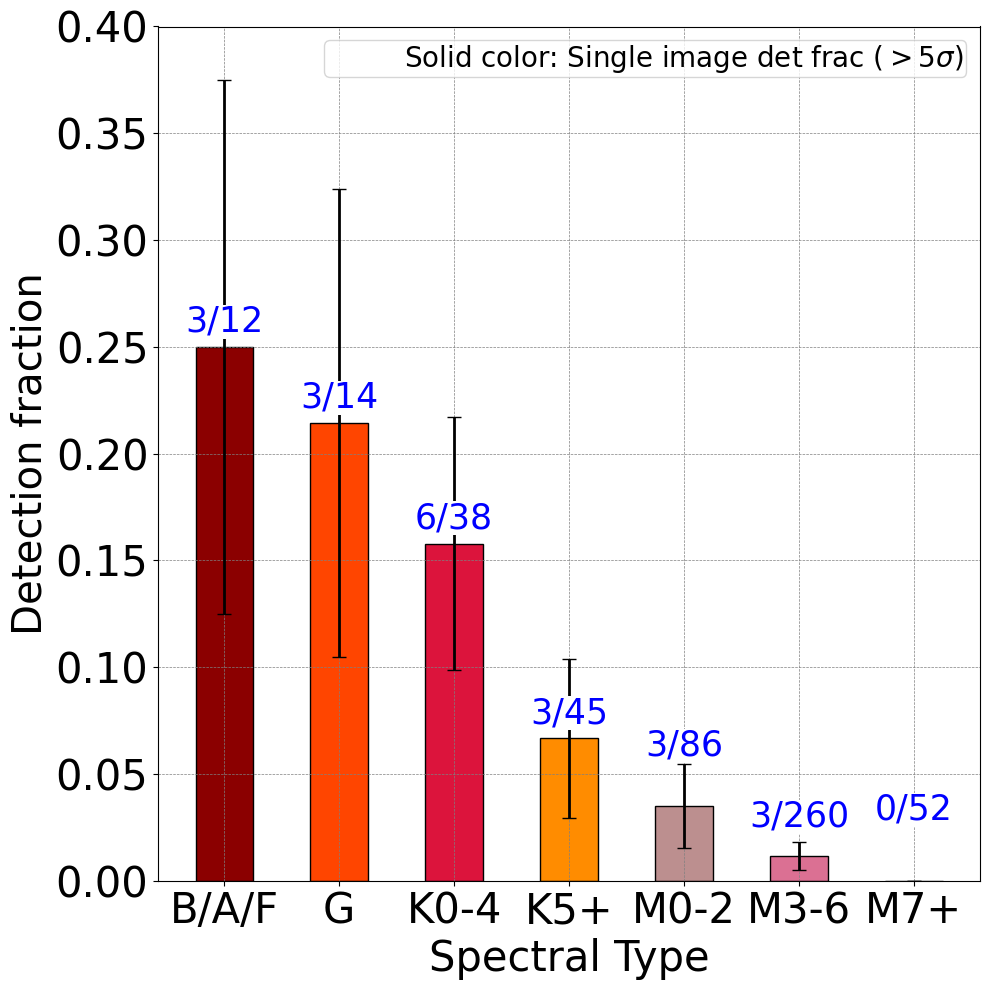}{0.48\textwidth}{(d)}}
\caption{(a) and (b) Plots of flux densities of sources detected in either or both Epoch 1 and 2 as a function of stellar spectral type, with upper limits represented by blue markers (i.e., stacked detections are not reflected in plots a or b). 
(c) Plot of the detection fraction (of sources detected at a ${>}$ 4$\sigma$ level) as a function of stellar spectral type, with these fractions noted above each bar, where stacked detections are also included. 
Hatches indicate the detection fractions for stacked detections.
(d) Plot of the detection fraction (of sources detected at a ${>}5\sigma$ level) as a function of stellar spectral type. 
} 
\label{fig:spt}
\end{figure}

\subsubsection{A radio luminosity - spectral type dependence}
Fig.~\ref{fig:spt} shows the flux density of the detected YSOs at epochs 1 and 2, and their relative detection fractions (i.e., the ratio of the number of detected YSOs (in \textit{either} epoch 1 or 2) within a specific spectral type bin to the total number of YSOs in that spectral type bin) all as a function of their spectral type. 
Of the 512 bona-fide Taurus members that we consider in our analysis, 507 have determined spectral types \citep[see][and Table 1 for all references]{Krolikowski2021} and thus our discussion of spectral types can be considered complete at the level of 99\% of the full sample.
Assessing the uncertainties attributed to these SpTs, 499/507 have quantified uncertainties (with a sample mean of 0.6, and a standard deviation of 0.6) and 8/507 have no uncertainties quantified (6 have SpTs in the range A3--B9, and 2 have SpTs in the range M0--M1).
Most prominent in this figure is the lower-left plot, which shows a trend between SpT and the detected fractions of YSOs, demonstrating higher detection fractions for earlier-type stars.
There is a monotonic increase in detection fractions from late to early spectral types, where we observe the following detection fractions: 25\% $\pm$ 13\% (B, A and F-types combined), 21\% $\pm$ 11\% (all G-types), 18.4\% $\pm$ 6.3\% ({K0--K4} types), 15.5\% $\pm$ 5.4\% (K5--K9 types), 7.0\%  $\pm$ 2.7\% (M0--M2 types), 2.3\% $\pm$ 0.9\% (M3--M6 types), and 1.9\% $\pm$ 1.9\% (SpTs later than M7).
To check the robustness of this monotonic trend, we also present in the lower-right plot detections at the ${>}5\sigma$ level.
For YSOs detected at the ${>}5\sigma$ level, there is a similar monotonic increase, where we observe the following detection fractions: 25\% $\pm$ 13\% (B, A and F-types combined), 21\% $\pm$ 11\% (all G-types), 15.4\% $\pm$ 5.9\% (early K-types), 6.7\% $\pm$ 3.7\% (late K-types), 3.5\% $\pm$ 2.0\% (early M-types), 1.2\% $\pm$ 0.7\% (late M-types).
We note that the uncertainties associated with the source SpTs are much smaller than the coarse bin sizes adopted to produce panels c) and d) in Fig.~\ref{fig:spt}. In addition, for the 8 sources without uncertainties in their SpT, 6 are members of the broadest B/A/F bin (which are highly unlikely to be erroneously classified), and 2 in the M0--2 bin (if either of these two sources were erroneously classified and were instead in either the K5+ or M3--6 bins, this would not alter the trend as seen). As such, the monotonic decrease in YSO detection fraction versus SpT bin appears robust, given the overall uncertainties associated with detection counts per SpT bin.

To examine if this result is skewed by the wide range of source distances present within the Taurus-Auriga YSO sample, we consider the luminosity densities of all sources ($L_R$), and present the values (for detections in cgs units) in Table~\ref{tab:xraytable} (to allow for direct comparison with these data in standard unit conventions), as well as the distance-independent flux densities ($F^{\prime}_{\rm{140pc}}$, in units of $\mu$Jy) for which
\begin{equation}
F^{\prime}_{\rm{140pc}} = F_{\nu} \times \Big[\frac{d}{140\,\rm{pc}}\Big]^2 = \frac{L_{R}}{4 \pi \times [140\,\rm{pc}] ^2},
\end{equation}
to simplify direct-comparisons between the source detections presented in this study and those in others, regardless of their distances.
This definition of $F^{\prime}_{\rm{140pc}}$ is directly proportional to the radio luminosity density, and thus any dependence on $F^{\prime}_{\rm{140pc}}$ is identical to any such dependence on radio luminosity density. We calculate two separate values for $F^{\prime}_{\rm{140pc}}$, one an epoch-averaged value (which represents one measure of the mean radio luminosity density of each source), and the other on a per-epoch basis (which represents the luminosity density of a source in any given VLASS observation, and thus does not average over sources that show significant variability). We describe the process to derive $F^{\prime}_{\rm{140pc}}$ in Appendix~\ref{appA1}. Although we have shown the presence of an SpT dependence considering the full sample, we next consider how this dependence is affected by binning up the luminosity density distributions.

\begin{figure}
    \centering
    \includegraphics[width=\textwidth, clip, trim={2cm 0cm 2cmcm 0cm}]{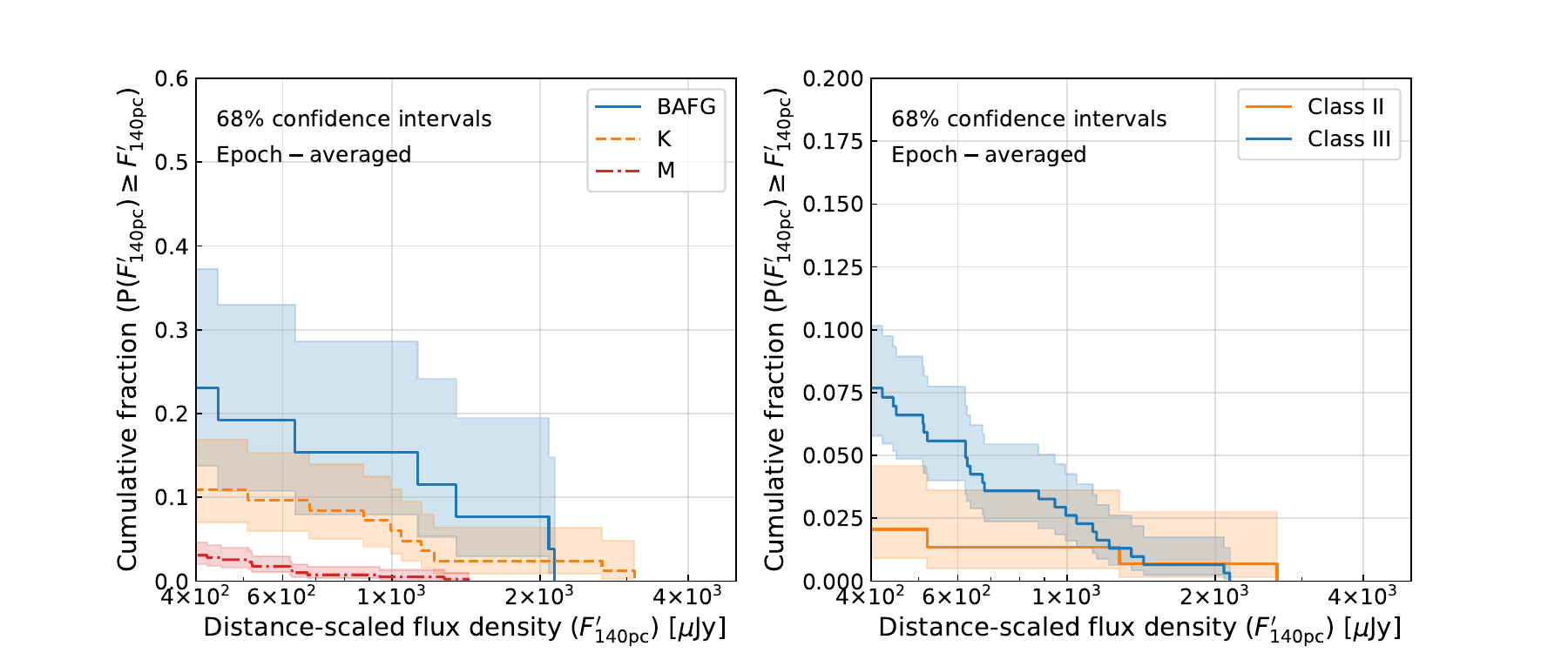}
    \includegraphics[width=\textwidth, clip, trim={2cm 0cm 2cmcm 0cm}]{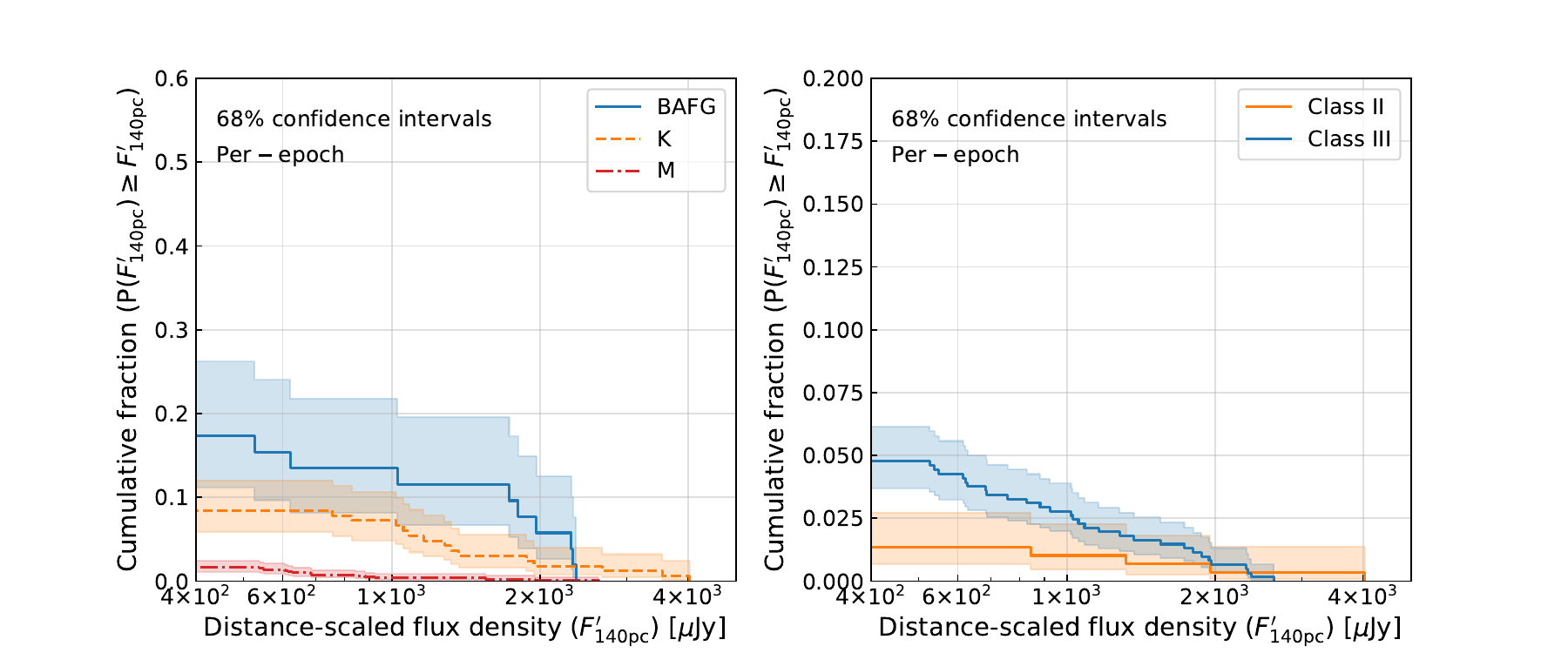}
    \caption{Cumulative density distribution plots of the distance-scaled flux densities ($F^{\prime}_{\rm{140pc}}$) for (left) all 507 sources based on their spectral types, and (right) for the 453 sources that are either class~II or class~III, with the top row representing data taken from average values of $F^{\prime}_{\rm{140pc}}$ over epochs 1 and 2, and the bottom row considering these data independently per-epoch.}
    \label{fig:LrCumulplots}
\end{figure}

We present on Fig.~\ref{fig:LrCumulplots} (left) the cumulative density distribution of $F^{\prime}_{\rm{140pc}}$ split into three coarse spectral type bins, chosen to combine the earliest, mid- and latest SpT stars, (top: for the epoch-averaged distribution, and bottom: for the per-epoch distribution). These coarser bins have detected population fractions of $23\%{\pm}8\%$ (6/26), $17\%{\pm}4\%$ (14/83) and $3.3\%{\pm}0.9\%$ (13/398) to more clearly present the radio luminosity density distribution of the full sample.
We produce these distributions using the {\tt lifelines} package \citep{Davidson-Pilon2019}, utilising a Kaplan-Meier estimator with `left-censoring' applied to all upper-limits \citep[advised as per][i.e., to ensure non-detections are statistically accounted for correctly]{Fiegelson1985}. 
The cumulative distributions demonstrate that the luminosity densities of M--type YSOs are significantly fainter than those of both K--type YSO populations and the B, A, F and G--type YSO populations. Moreover, whilst the cumulative fractions are most different over the complete sample (i.e., in the very first sample bins) the M-type distribution is lower than the other distributions over approximately all luminosity density ranges (either per-epoch, or epoch-averaged).
The luminosity density distributions are not significantly different between the K--type and early SpT bins (which may instead reflect a steep drop in typical luminosities between K and M-type sources). We note there remains a skew that appears to favor the earliest SpT sources having a luminosity density enhancement. More sensitive data is needed to understand if the underlying luminosity density distribution falls monotonically across the full sample, or if the distribution is bimodal.
In combination with the enhancement in the detection fractions of earlier type YSOs, and their enhancement in their luminosity density distributions, we thus present evidence of a radio luminosity dependence on YSO spectral type in Taurus--Auriga.

\subsubsection{Interpreting the spectral type dependence} \label{sec:SpTInterpret}
Whilst previous surveys of Taurus were much higher sensitivity than available with VLASS data, such as that of \citet{Dzib2015} which covered 127 target fields across the region, our study examines $\sim$5 times as many source locations. This broader coverage enables us to fully sample the Taurus-Auriga star-forming region using VLASS data. In this work, our ability to measure the radio luminosity density distribution in an unbiased manner for all stars in the Taurus-Auriga star forming region has recovered a dependence on spectral type, which to our knowledge has not been reported before in Taurus (nor elsewhere).
{Our finding is particularly interesting given the results presented by \citet{Launhardt22}, where a sample of M5--F0 \textit{main sequence} stars are shown to have a monotonically \textit{falling} radio detection fraction with SpT (from 50\% down to 10\%).
The trend observed for main sequence stars is reversed in the VLASS data, suggesting a possible fundamental difference between the radio luminosity distributions of YSOs and main sequence sources.

The origin of the observed dependence for YSOs nevertheless remains unclear. Two limitations of this study are that VLASS Quick Look images provide data at i) a single frequency band, and ii) with Stokes I (total intensity), and thus cannot robustly measure radio spectral index nor polarization that can be used to infer underlying radio emission mechanisms \citep{Gudel02}. Thus, we cannot yet determine which physical processes are responsible for the observed emission.

Moreover, with just two time bins per epoch, we remain limited in determining which detections are from quiescent emission or from short-term flaring/long-term variability. 
For 7/28 detections we have assessed the emission to be variable, and thus plausibly associated with stellar flares.
Although these sources are clearly variable, we cannot robustly infer the quiescent levels of their radio emission with two time bins. For the remaining detected sources, we cannot rule out the possibility that they too are flaring at lower levels; for example, the faintest detected sources are consistent with being variable at the level of ${\lesssim}100$\%, falling to ${\lesssim}30$\% for the brighter sources (either of which could represent substantial changes in radio luminosity). 
Flaring can be driven by intrinsic stellar processes (e.g., stellar rotation) and external processes (e.g., interactions with nearby companions), thus the presence of 3\,GHz radio emission from flaring may too have inherent trends with stellar properties and be primarily responsible for the luminosity--SpT dependence.

Our main focus in this work is to understand the typical radio properties of a complete sample of YSOs, which is likely driven by a combination of flaring and quiescent emission at 3\,GHz.
As such, here we consider two plausible scenarios that may explain the SpT dependence, in the context of the relatively high fraction of variable emission sources and stellar multiples in the sample.

\textit{{Stellar multiplicity?}}} 
The VLASS data implies that the typical radio luminosity of an earlier-type YSO is enhanced versus a later-type YSO.
This dependence may reflect the intrinsically higher stellar multiplicity fraction of earlier type YSOs/intermediate-mass stars in comparison to lower-mass stars \citep{Duchene13}.
From our analysis of stellar multiplicity, at least 39\% of the detected sources are in stellar binaries/multiples, and this is likely a lower-limit based on the expected multiplicity fraction of Taurus-Auriga YSOs (see \S\ref{sec:binariesDets}).
Additional emission from unresolved companions, or from stellar binary/multiple interactions could therefore increase radio emission towards earlier-type YSOs \citep[and induce variability at the levels seen in the full sample, given well-known examples of strong (periodic) radio/millimeter luminosity enhancements from e.g., DQ~Tau and V773~Tau, see][respectively]{Massi06, Salter10}.
To test this hypothesis, we suggest i) further long-term radio monitoring of the detected sample to investigate the occurrence of periodic flaring from binary interactions and ii) high-resolution follow-up sufficient to resolve multiple sources contributing to the observed radio emission.

{\textit{{Magnetic activity?}}}
The full sample is entirely comprised of YSOs, known to be rapid rotators. An alternative explanation for the observed radio emission is intrinsic processes of young stars, e.g., magnetic reconnection events which can induce significant levels of quiescent radio emission and strong flaring events, even in the absence of stellar companions or interactions with any such companions. In this scenario, the radio luminosity distribution (for sources dominated by quiescent emission) would arise from magnetic activity and follow the classical G\"udel-Benz Relationship \citep[herein GBR;][which we detail further in Appendix~\ref{appA}]{Benz93, Benz94}. We test the validity of this scenario by considering whether these are consistent with the GBR (and being magnetically active) by plotting their location on the $L_X$--$L_R$ plane (excluding the 7 variable sources), see Fig.~\ref{fig:GBR}. We find that all sources (except class~II HK~Tau~A) obey the GBR, leaving open the possibility that the sample dependence on SpT originates from earlier-type YSOs being typically more magnetically active than later-type YSOs. 
For example, this dependence (and the systematically lower detection fractions of later type stars) may reflect that these stars are intrinsically fainter (in terms of their bolometric and X-ray luminosities), which manifest in being less luminous at radio wavelengths (and in particular for the shallow sensitivities provided by VLASS).
To test this hypothesis, we suggest deeper radio and X-ray observations of the full sample to better investigate the connection between their X-ray and radio properties.

\subsection{Radio luminosities as a function of evolutionary YSO class}
We note that \citet{Krolikowski2021} do not present YSO mid-infrared classifications of our sources. 
We generate these from the the allWISE catalog \citep{Cutri2013}. 
We cross-match the Taurus-Auriga catalog \citep{Krolikowski2021} with the allWISE catalog, which identified 551/587 source matches.
We exclude all sources that do not meet our photometric quality flag (`qph') criteria (SNR${>}3$, i.e. qph values of either `A' or `B') for both the K-band and W3-band data in the WISE data \citep[K-band \textit{2MASS} data is incorporated into the allWISE catalog][]{Skrutskie06}.
We calculate mid-infrared spectral indices for all sources, adopting $\alpha_{\rm{2-12\mu m}}$ \citep[i.e., from $\alpha = \partial \log{\nu F_\nu} / \partial \log{\nu}$, for 2.2 and 12\,$\mu$m e.g.,][]{Williams2011}.
From this same equation we translated the spectral slope $\alpha_{\rm{2-12\mu m}}$ into a [K]--[12] color constraint, for which [K]--[12]${<}$2.19 corresponds to class III, 2.19${<}$[K]--[12]${<}$4.61 corresponds to class II, and [K]--[12]${>}$4.61 includes all class 0s, Is and Fs.
On manual inspection, we found a number of well-known IRAS sources and class~III stars to have WISE photometric flags suggesting these were `confused' and for completeness we included these in our sample bins. 
In total, we assign YSO classes to 478/512 of the \citet{Krolikowski2021} bona-fide Taurus-Auriga YSOs, i.e., the full sample is complete to the level of 93\% (478/512), with 25 class~0/I/F sources (5.2\%), 147 class~II sources (30.8\%), and 306 class~III sources (64.0\%). The relative splits in class broadly agree with previous determinations in Taurus. For example, comparing with those of \citet{Esplin19}, we find those have fractions at the level of 8\%, 38\%, and 54\% for the same combined class~0/I/F, class~II, and class~III bins that we adopt here. The absolute differences between the fractional levels between the determination here and those of previous works are most plausibly due to the different stellar samples of \citet{Krolikowski2021} and \citet{Esplin19}, e.g., since 108 sources (of the 512 YSOs) present in \citet{Krolikowski2021} are not included in the table of \citet{Esplin19}.


\begin{figure}[th!]
\gridline{\fig{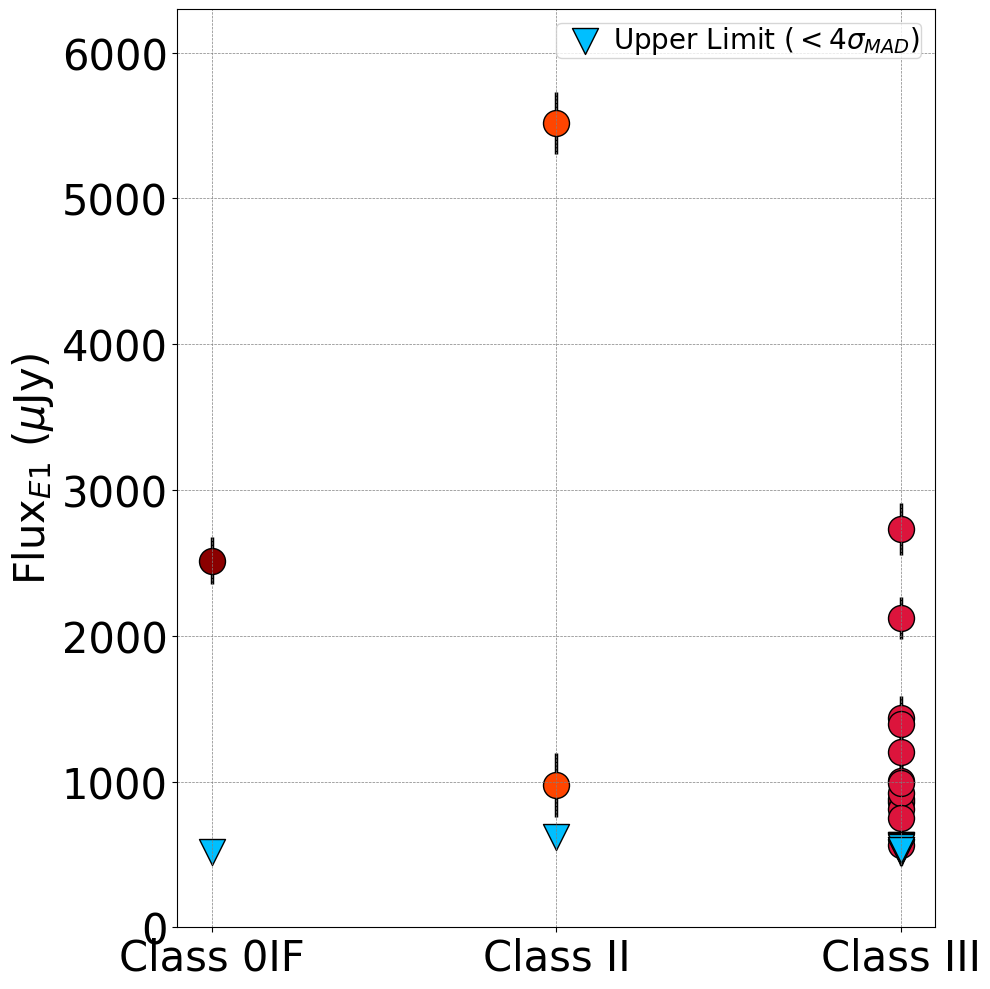}{0.5\textwidth}{(a)}
          \fig{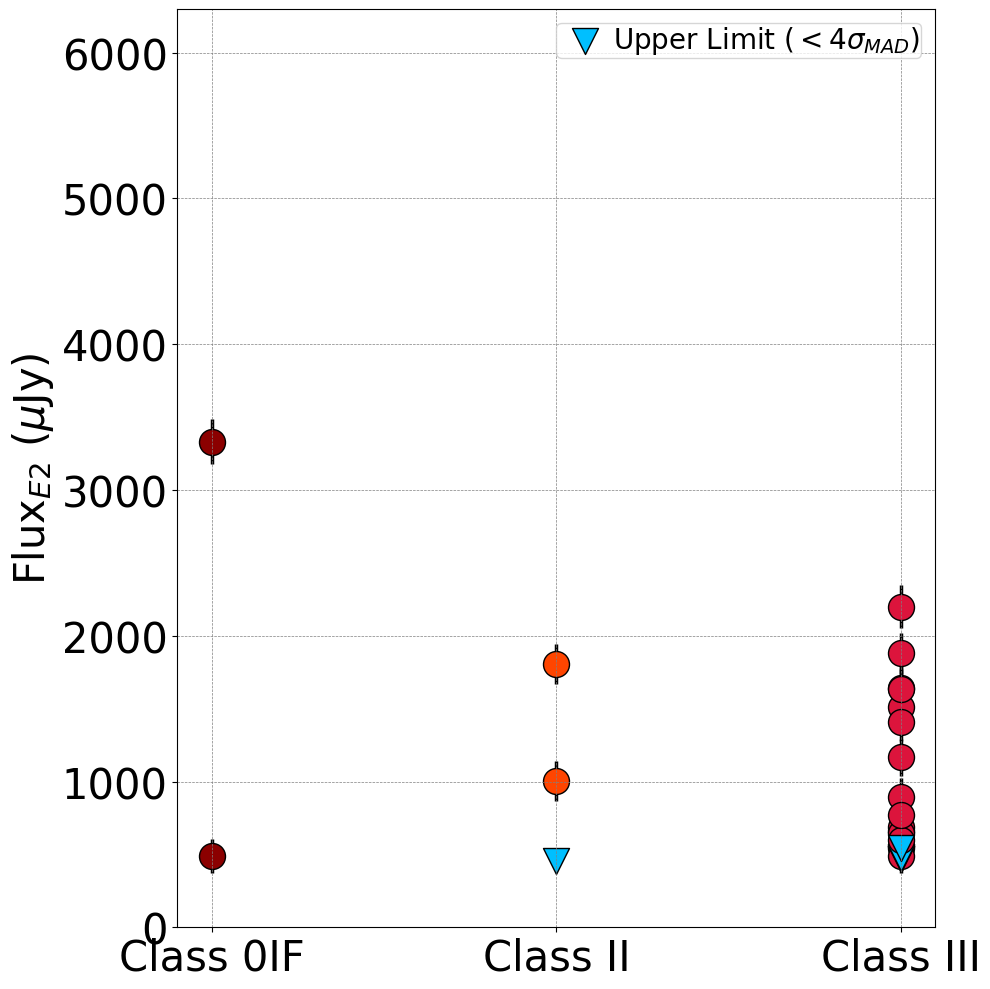}{0.5\textwidth}{(b)}}
\gridline{\fig{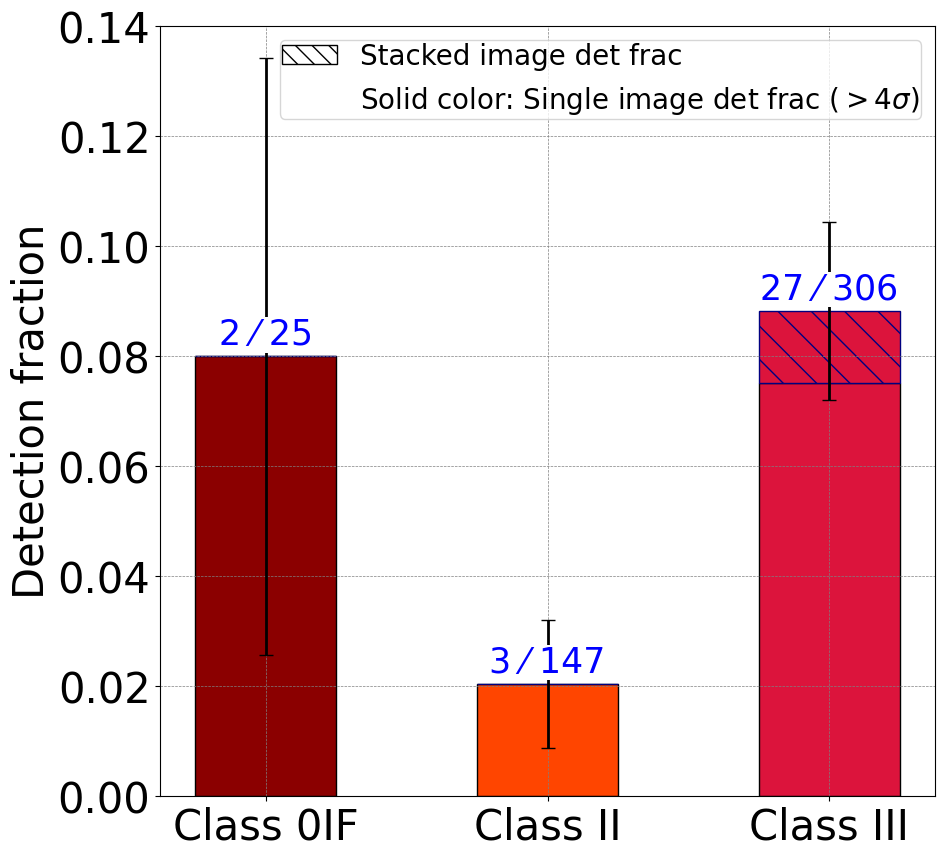}{0.5\textwidth}{(c)}
          \fig{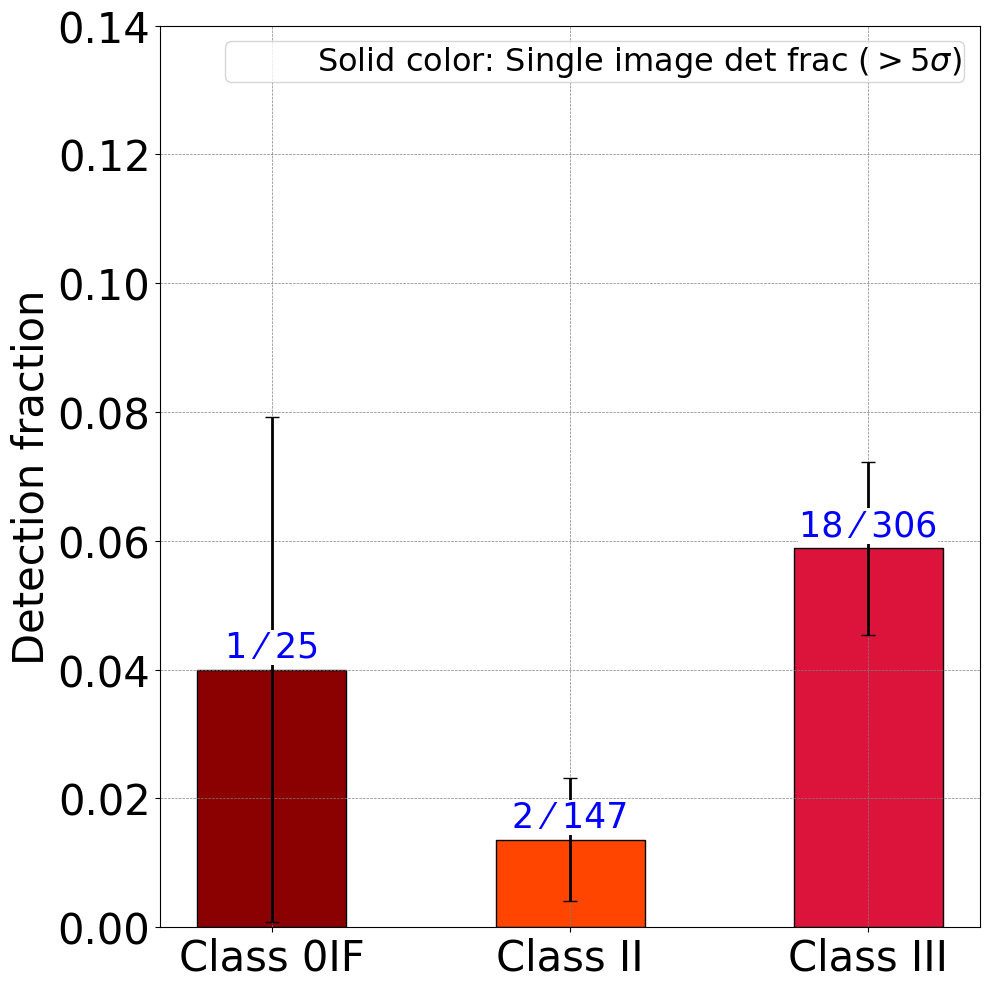}{0.45\textwidth}{(d)}}
\caption{(a) and (b) Plots of flux densities of sources detected in either or both Epoch 1 and 2 as functions of YSO class, with blue markers indicating upper limits (i.e., stacked detections are not reflected in plots a or b). 
(c) Plot of the detection fraction (of sources detected at a ${>}4\sigma$ level) as a function of YSO class, which includes stacked detections. 
Hatches indicate the detection fractions for stacked detections. 
Since we cannot determine the YSO class of one of our 33 detected YSOs, only 32 of our detections are reflected on plot c.
(d) Plot of detection fraction (of sources detected at a ${>}5\sigma$ level) as a function of YSO class.}
\label{fig:class}
\end{figure}

Fig.~\ref{fig:class} shows the flux density of the detected YSOs at epochs 1 and 2, and their relative detection fractions (in either epochs 1 or 2) as a function of their respective YSO class bins.
There is little correlation between the absolute flux levels and YSO class (plots a and b), but instead that the rate of detections significantly alters with YSO class (plots c and d).
We see towards the earliest stage class~0/I/F YSOs a detection fraction of 8.0\% $\pm$ 5.4\% (2/25), a detection fraction towards the class~II YSOs of 2.0\% $\pm$ 1.2\% (3/147), and towards the class~IIIs a higher fraction of 8.8\% $\pm$ 1.6\% (27/306). 
To check the robustness of this change in fractional levels, we also present in the lower-right plot detections at the ${>}5\sigma$ level.
For YSOs detected at ${>}5\sigma$, we see towards class~0/I/F YSOs a detection fraction of 4.0\% $\pm$ 3.9\%, a detection fraction towards the class~II YSOs of 1.4\% $\pm$ $1.0\%$, and towards the class~IIIs also a higher fraction of 5.9\% $\pm$ 1.3\%.
Considering the detection fractions of plot c, the class~0/I/F and either the class~II detection fraction or the class~III detection fractions are broadly consistent to within $2\sigma$ (i.e., considering binomial statistics for the detection fractions in the two populations).
In contrast, the increase in the detection fraction from class~II to III is significant (i.e., the difference in the fractional detection rates of the class~II and III bins divided by their combined uncertainties in quadrature is $3.4\sigma$).
We note that a similar trend was found by \citet{Dzib2015} in Taurus, where the mean flux densities of class~III YSOs were shown to be higher than class~II YSOs (at both 4.5\,GHz and 7.5\,GHz). A like-for-like comparison between \citet{Dzib2015} and the results presented here is challenged by the lower frequency and sensitivity of VLASS data, however these appear to indicate a similarity in the flux distribution as a function of YSO class.


As earlier, to examine if this result is skewed by the wide range of source distances present within the Taurus-Auriga YSO sample, we investigate whether this trend between YSO class and radio brightness is reflected in the radio luminosity density distributions (utilizing the same Kaplan-Meier estimator with `left-censoring' applied to all upper-limits as used in \S\ref{sec:LR_spt}, instead binning by YSO class).
We present on Fig.~\ref{fig:LrCumulplots} (right) the cumulative density distribution of $F^{\prime}_{\rm{140pc}}$ split by YSO class (only presenting the class~II and III distributions given the small sample size of the class~0/I/F YSOs, likewise providing separate values for $F^{\prime}_{\rm{140pc}}$ based on an epoch-averaged value, and the other on a per-epoch basis. 
The cumulative distributions demonstrate these distributions to be distinct over the complete sample of class~II and III YSOs (i.e., in the first bins) with the class~IIIs constituting a greater fraction of the detected sources. These distributions become consistent for bins including only the brightest luminosity density sources (though these bins contain few sources).
As such, we therefore infer this increase in detection fraction at the class~III stage to reflect an enhancement in detectable YSO radio emission as these young stars evolve.

\subsubsection{YSO class completeness}
The relative completeness level of the sample for YSO class is high at 93.3\% (478/512).
We note that the majority of stars without reliable 2MASS or WISE data are more likely to be class~III stars (given these would be significantly brighter at near- and mid-infrared wavelengths, and so would have been detected as \textit{IRAS}/\textit{Akari} sources if these were early-stage YSOs).
Accounting for this YSO class bias in the completion levels, we note that the lowest possible class~III detection rate would be 7.9\% $\pm$ 1.5\% (27/340), e.g., assuming all 34 (512 minus 478) stars not included in our analysis all happen to be class~III.
In this instance the difference in detection fractions between the class~II and III stages remains significant (at the $3.1\sigma$ level), thus demonstrating an evolutionary trend in these VLASS detections.

\subsubsection{Interpreting the YSO class dependence}
As discussed earlier, without data to measure the radio spectral slopes or polarized emission levels, our interpretations for the underlying cause of the YSO class dependence are limited. 
Nevertheless, there are well-known physical differences between the radio properties of different class YSOs that likely contribute to the VLASS data. For example,
at the class~0/I/F stage, stars are readily accreting and are known to host strong winds and dust-rich disks, both of which contribute to their radio emission \citep[see e.g.,][]{Schwartz84, Rodmann06, Pascucci12}.
At the class~II stage, stellar accretion processes and winds are still in operation (albeit with lower strengths than earlier class YSOs) and circumstellar dust is present (albeit with lower masses than earlier class YSOs), which contribute to their radio emission, though perhaps more weakly. 
Finally, at the class~III stage, wind-losses and accretion have ended, and circumstellar disks have near/fully dispersed \citep[see e.g.,][]{Lovell2021a,Michel21}, however likely owing to their ongoing rapid rotation, their magnetic field strengths give rise to powerful flares, and thus radio-bright stellar emission \citep{Andre1992}.

Whatever the underlying radio emission mechanism at play, the VLASS data suggest that the radio luminosities of YSOs at these distinct YSO classes typically span the same ranges \citep[likewise noted by ][]{Dzib2015}.
Extending the radio analysis of the Taurus-Auriga star-forming region to a full sample of YSOs, we show marked differences in the detectability of YSOs at different stages.
Whilst the absolute radio brightnesses of class~II YSOs can match those of class~0/I/F and III YSOs, these are observed less frequently, and thus overall, these may represent a fainter radio \textit{population}.
One possible interpretation of this trend is that the evolution of the circumstellar disk, stellar accretion and outflows have a strong impact on the evolution of YSO radio luminosities (and since the VLASS data only show evolutionary differences between the class~II to III stages, we remain agnostic to evolution between the earlier stages). 
For example, whilst class~II YSOs host optically thick outer material which may obscure bright stellar processes, by the class~III stage, any outer optically thick material has dispersed, which allows radio emission closer to the stellar surface to radiate outwards unimpeded. Lacking an understanding of the radio emission mechanisms in operation for the full sample, we cannot test this hypothesis. However, these VLASS data nevertheless argue that the typical radio emission of class~II YSOs is (on average) fainter than that of class~III YSOs, and thus the \textit{typical} luminosity of stellar processes from YSOs may exceed that from winds/disks of class~II YSOs. Additional observations are required to advance any further understanding of this conclusion.

\subsection{Other correlations?}
This survey results present us with the opportunity to investigate trends within the sample, in terms of their radio fluxes, luminosities, and detection fractions, in the context of the \citet{Krolikowski2021} study of the Taurus star forming region.
We find no significant trend between radio detection fractions and the \citet{Krolikowski2021} sub-group IDs (noted `GMM' (Gaussian mixture model) in Table~\ref{tab:detections}, which define 17 stellar sub-groups within Taurus--Auriga).
We measure these detection fractions to range from 0--17\%, with a mean of 5\%. 
Moreover, by binning all sub-group members denoted by \citet{Krolikowski2021} as either in `C' (compact) sub-populations or `D' (distributed) sub-populations, we find detection fractions of 6.3\% $\pm$ 1.4\% (20/317) and 6.7\% $\pm$ 1.8\% (13/195) respectively.
As such, we see no dependence on the radio luminosity distribution between different GMM sub-regions, nor the distributed or compact categories.

We find no significant trend between radio detection fractions and the stellar ages defined by \citet{Krolikowski2021}, for the three age bins ${<}2$\,Myr, $2-3$\,Myr and ${>}3$\,Myr.
We measure detection fractions of 5.3\% $\pm$ 1.7\% (9/171), 6.7\% $\pm$ 1.8\% (13/195) and 7.7\% $\pm$ 2.2\% (11/142) in these age bins respectively. Though there is a strong correlation between YSO ages and classes, the lack of an observed age trend with radio luminosity does not contradict the trend present with radio luminosity and YSO class. While the overall fraction of protoplanetary disks decreases with age across star-forming regions, the variance in the dispersal timescales of protoplanetary disks is several Myr \citep[see e.g.,][]{Evans2009,Michel21,Monsch2023}. This substantial spread in disk dispersal times coupled with the coarse choice of age bins results in class~III YSOs being the dominant YSO class (and contributor to each binned detection fraction) in each age bin.
Overall, our analysis with VLASS provides no evidence for radio luminosity evolution between these ages.

\section{Conclusions and Summary} \label{sec:conclusions}
Here we have presented RADIOHEAD: The Radiowave Hunt for young stellar object Emission And Demographics, in which we measured the radio continuum fluxes and luminosities with public VLA Sky Survey (VLASS; at 2--4\,GHz, with $\sigma_{\rm{VLASS}}{\sim}110-140\,\mu$Jy, $2.5''$ resolution) at all of the locations of {\it Gaia}-confirmed members of the Taurus star forming region \citep[as defined within][]{Krolikowski2021}.
We measured significant bright radio emission (${>}4\sigma$) for 35 sources coincident with the locations of \citet{Krolikowski2021} sources, of which 33 are associated with Taurus-Auriga YSOs, spanning 2 complete VLASS epochs (2019 and 2021), and the partially observed 2023 epoch.
Of these 33 detections, 28 are observed in single epoch images, whereas 5 are obtained by stacking two separate images. Overall, this work demonstrates that YSOs are detectable with VLASS, providing new avenues to explore the radio luminosity of a near-complete YSO population across the VLASS footprint. Though YSO radio emission from even nearby YSOs is faint (i.e., 6.0\% of all source locations yielded a detection), VLASS's complete coverage of the northern sky with \textit{Gaia}-confirmed sources enables novel population-based studies.

Between VLASS epochs 1 and 2, we measured 25\% (7/28) of our detected sources to have significant radio variability, ranging from a few tens of percent, to lower-limits of a few hundred percent. Our relatively high detection fraction of variable sources indicates a large fraction are likely from stellar flares. We found that 39.4\% $\pm$ 8.5\% (13/33) of detections appear around stars associated as binaries, a tentative enhancement in comparison to the binary fraction of the sample (18.5\% $\pm$ 1.7), but note this may simply reflect a systematic bias in the quantified levels of binaries stated in \citet{Krolikowski2021}.

We found a correlation between the detection rates of YSOs with stellar spectral type, wherein the fractional detection rate of radio emission coincident with earlier type stars is systematically higher than later type stars.
For all sources we measure their radio luminosity densities, and present cumulative density distributions of radio luminosity densities that further demonstrate significant luminosity density enhancements for early versus late type stars.
We discuss this enhancement in the context of earlier type stars being more typically hosted in binaries as one plausible explanation to the observations.

We found that mid-infrared YSO class to be an indicator of radio detectability, with class~III stars being detected at a higher rate versus class IIs.
We found no correlation between radio emission and either the ages of YSOs, or their spatial locations, indicating insignificant evolution in the radio luminosity distribution over the few Myr age span of the bulk of Taurus YSO ages.


\section{Software and third party data repository citations} \label{sec:cite}
The VLA data used here are from the VLASS project, epochs VLASS `1.1', `1.2', `2.1', `2.2' and `3.1', and were collected from the Canadian Astronomy Data Centre (CADC).
The National Radio Astronomy Observatory (NRAO) is a facility of the National Science Foundation (NSF) operated under cooperative agreement by Associated Universities, Inc. 
CIRADA is funded by a grant from the Canada Foundation for Innovation 2017 Innovation Fund (Project 35999), as well as by the Provinces of Ontario, British Columbia, Alberta, Manitoba and Quebec.
This research has made use of the NASA/IPAC Infrared Science Archive, which is funded by the National Aeronautics and Space Administration and operated by the California Institute of Technology.

\begin{acknowledgments}

We would first like to thank the anonymous referee for their careful and thorough review of our work, which has substantially improved its quality.
We extend our thanks to the NRAO helpdesk for assisting in understanding the relative independence of different VLASS image cutouts.
We thank Vinay Kashyap for discussions regarding our observed spectral type dependence.
R. A. Rahman acknowledges the SAO REU program for supporting this work. The SAO REU program is funded in part by the National Science Foundation REU and Department of Defense ASSURE programs under NSF Grant no. AST-2050813, and by the Smithsonian Institution.
J. B. Lovell and E.W. Koch acknowledge the Smithsonian Institute for funding via a Submillimeter Array (SMA) Fellowship.
E.W. Koch acknowledges support from the Natural Sciences and Engineering Research Council of Canada.
KM was supported by NASA {\it Chandra} grants GO8-19015X, TM9-20001X, GO7-18017X, HST-GO-15326 \& JWST-GO-1905.


\end{acknowledgments}

%

\vspace{5mm}
\facilities{Karl G. Jansky Very Large Array (VLA), Planck}


\software{astropy \citep{2013A&A...558A..33A,2018AJ....156..123A,AstropyCollaboration2022ApJ...935..167A},
          CASA \citep{CASA},
          TOPCAT \citep{TOPCAT}}



\appendix
\section{VLASS `Quick Look' images of detections} \label{appA}
\subsection{Detections in single images}
\begin{figure}[!htbp]
    \centering
    \begin{minipage}{1.0\textwidth}
    \centering
    \includegraphics[width=0.9\textwidth]{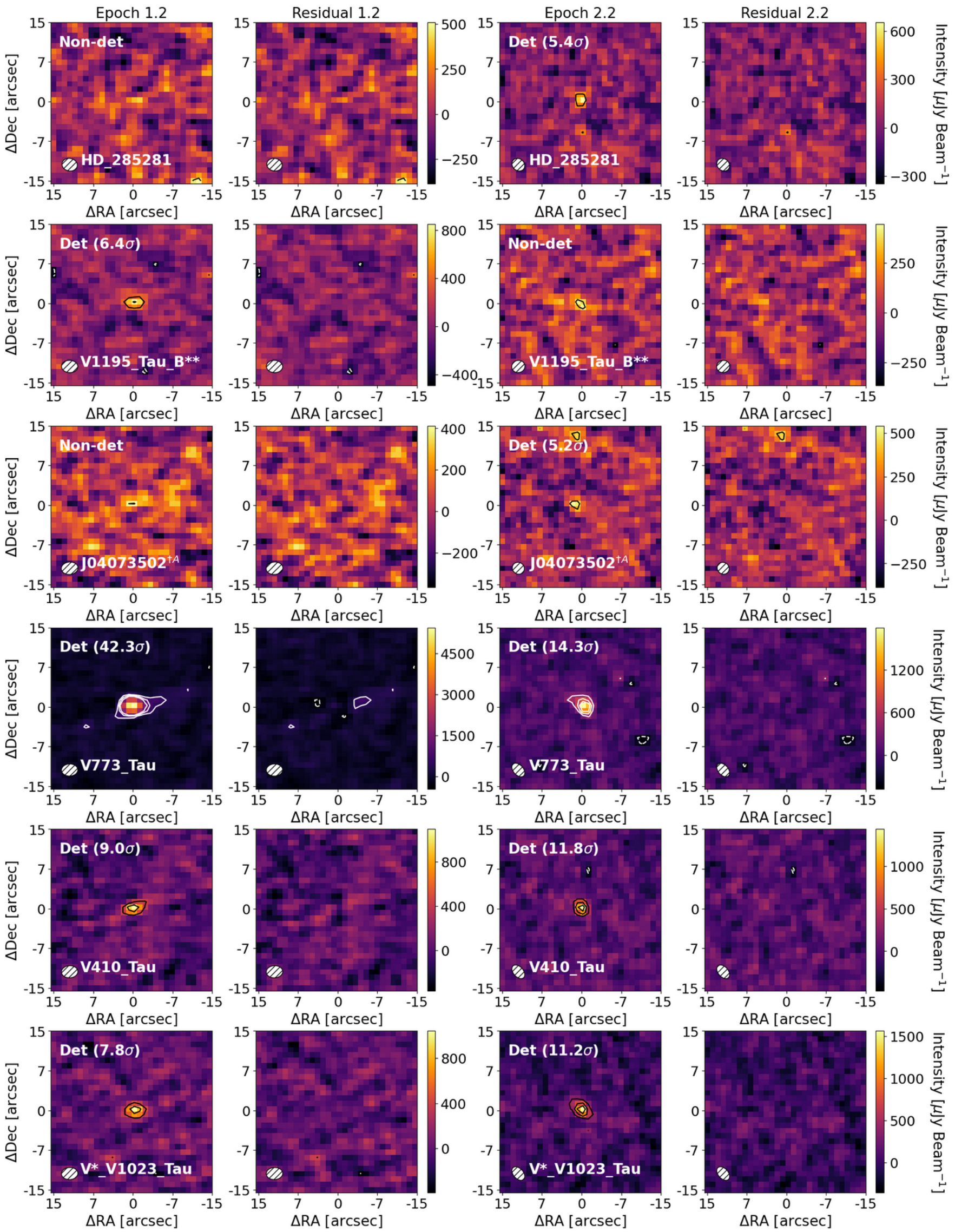}
        \caption{Single VLASS Quick Look Epoch 1 and Epoch 2 images of detections showing HD~285281, V1195~Tau~B, V1195~Tau~B, J04083502, V773~Tau, and V410~Tau as well as their residual maps. VLA beams are indicated in the lower left corners, while source SNR (for images with a detection) and non-detections are labeled on the upper left. Contours are set at [-3,3,6,9]$\sigma$ levels.}
        \label{fig:single_det_1}
    \end{minipage}
 \end{figure}

\begin{figure}[!htbp]
    \centering
    \begin{minipage}{1.0\textwidth}
    \centering
    \includegraphics[width=0.9\textwidth]{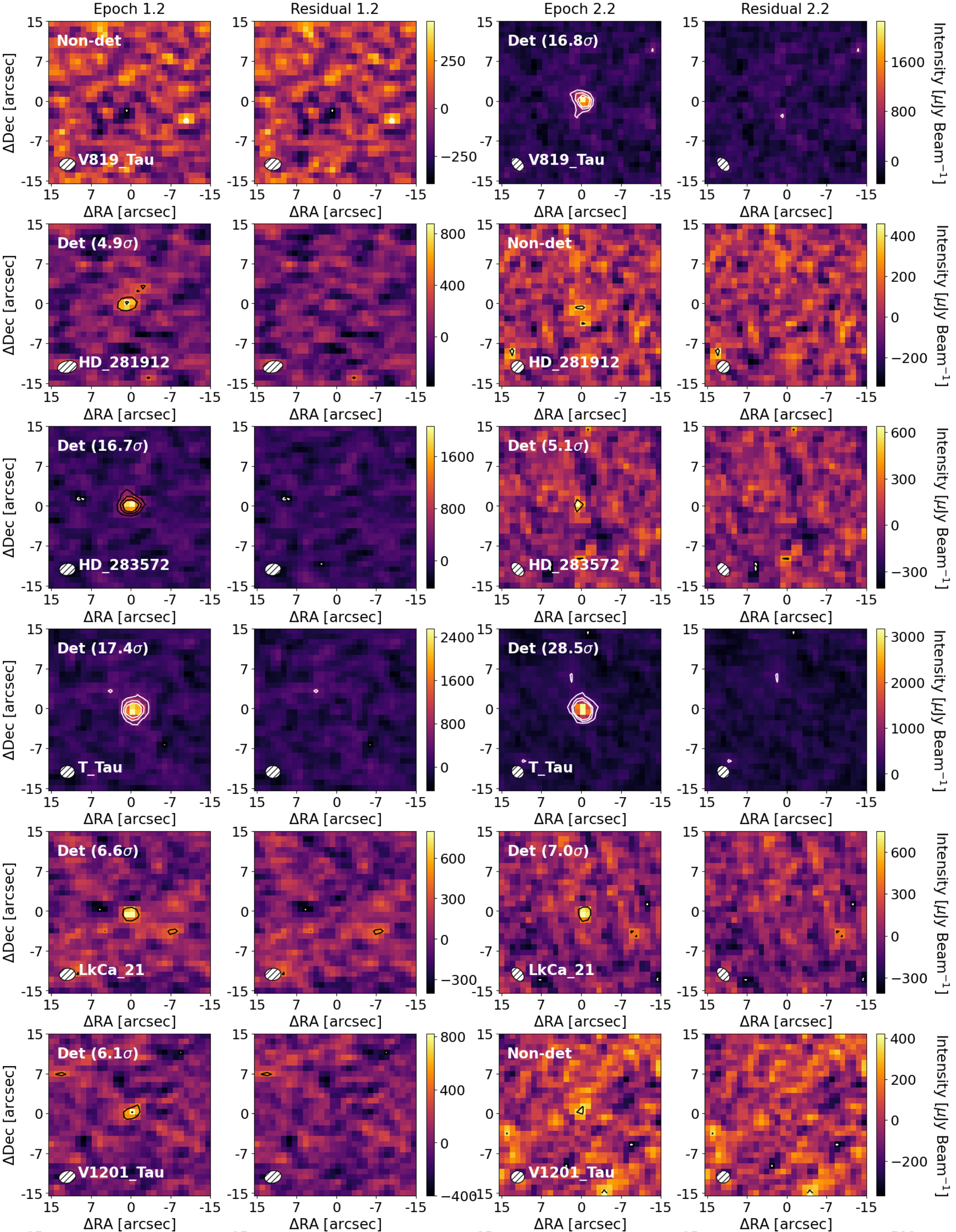}
        \caption{Single VLASS Quick Look Epoch 1 and Epoch 2 images of detections showing V819~Tau, HD~281912, HD~283572, T~Tau, LkCa~21, and V1201~Tau, as well as their residual maps. VLA beams are indicated in the lower left corners, while source SNR (for images with a detection) and non-detections are labeled on the upper left. Contours are set at [-3,3,6,9]$\sigma$ levels.}
        \label{fig:single_det_2}
    \end{minipage}
 \end{figure}

\begin{figure}[!htbp]
    \centering
    \begin{minipage}{1.0\textwidth}
    \centering
    \includegraphics[width=0.9\textwidth]{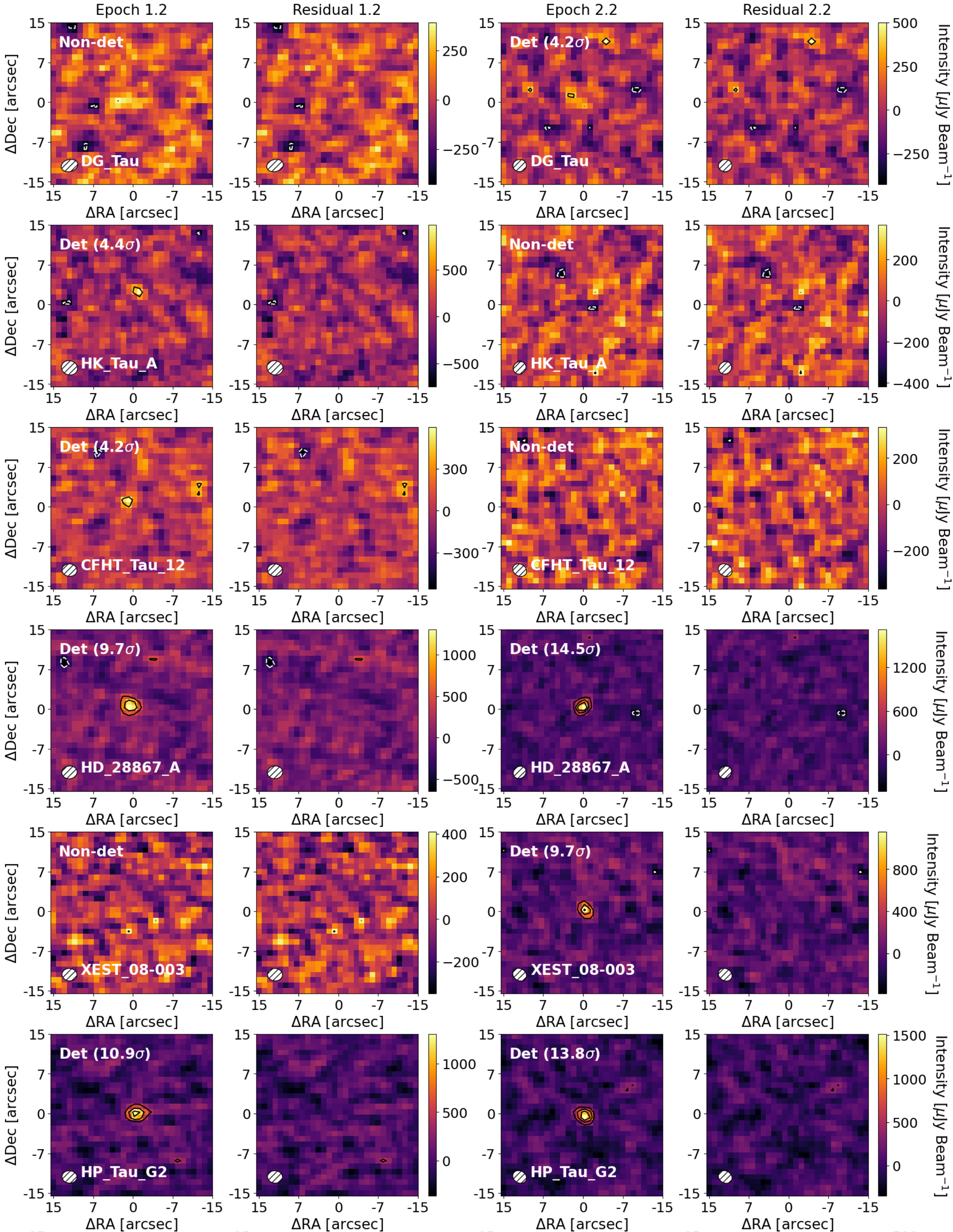}
        \caption{Single VLASS Quick Look Epoch 1 and Epoch 2 images of detections showing DG~Tau, HK~Tau, CFHT~Tau~12, HD~28867~A, XEST~08-003, and HP~Tau~G2, as well as their residual maps. VLA beams are indicated in the lower left corners, while source SNR (for images with a detection) and non-detections are labeled on the upper left. Contours are set at [-3,3,6,9]$\sigma$ levels.}
        \label{fig:single_det_3}
    \end{minipage}
\end{figure}

\begin{figure}[!htbp]
    \centering
    \begin{minipage}{1.0\textwidth}
    \centering
    \includegraphics[width=0.9\textwidth]{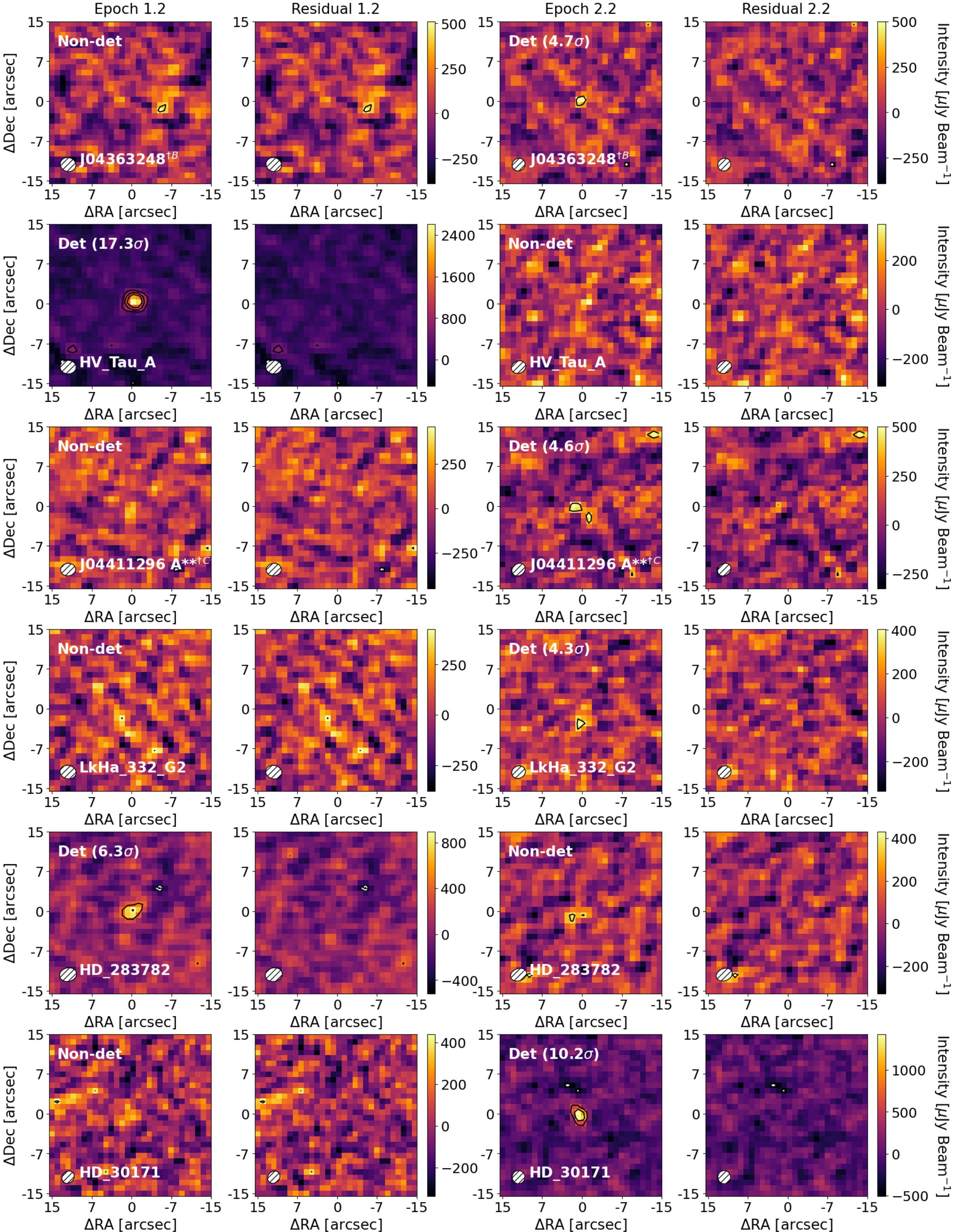}
        \caption{Single VLASS Quick Look Epoch 1 and Epoch 2 images of detections showing J04363248$^\dag$$^B$ (see caption of Table \ref{tab:detections}), HV~Tau~A, J04411296~A**$^{\dag}$$^{C}$ (see caption of Table \ref{tab:detections}, LkHA~332~G2, HD~283782, and HD~30171, as well as their residual maps. VLA beams are indicated in the lower left corners, while source SNR (for images with a detection) and non-detections are labeled on the upper left. Contours are set at [-3,3,6,9]$\sigma$ levels.}
        \label{fig:single_det_4}
    \end{minipage}
 \end{figure}

\begin{figure}[!htbp]
    \centering
    \begin{minipage}{1.0\textwidth}
    \centering
    \includegraphics[width=0.9\textwidth]{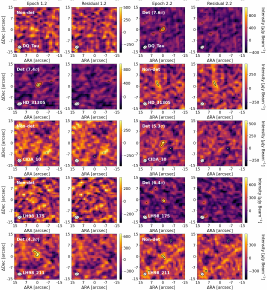}
        \caption{Single VLASS Quick Look Epoch 1 and Epoch 2 images of detections showing DQ~Tau, HD~31305, CIDA~10, LH98~175, and LH98~211, as well as their residual maps. VLA beams are indicated in the lower left corners, while source SNR (for images with a detection) and non-detections are labeled on the upper left. Contours are set at [-3,3,6,9]$\sigma$ levels.}
        \label{fig:single_det_5}
    \end{minipage}
 \end{figure}

Here we present images and residual images (following 2D model subtraction) for all 29 sources that presented detections in (at least) one single image in Figures \ref{fig:single_det_1}, \ref{fig:single_det_2}, \ref{fig:single_det_3}, \ref{fig:single_det_4}, and \ref{fig:single_det_5}.

\subsection{Detections in stacked images}
Here we present images and residual images (following 2D model subtraction) for all 6 sources that presented detections in their stacked images in Figure \ref{fig:stacked_dets}.
\begin{figure}
    \centering
    \begin{minipage}{1.0\textwidth}
    \centering
    \includegraphics[width=0.5\textwidth]{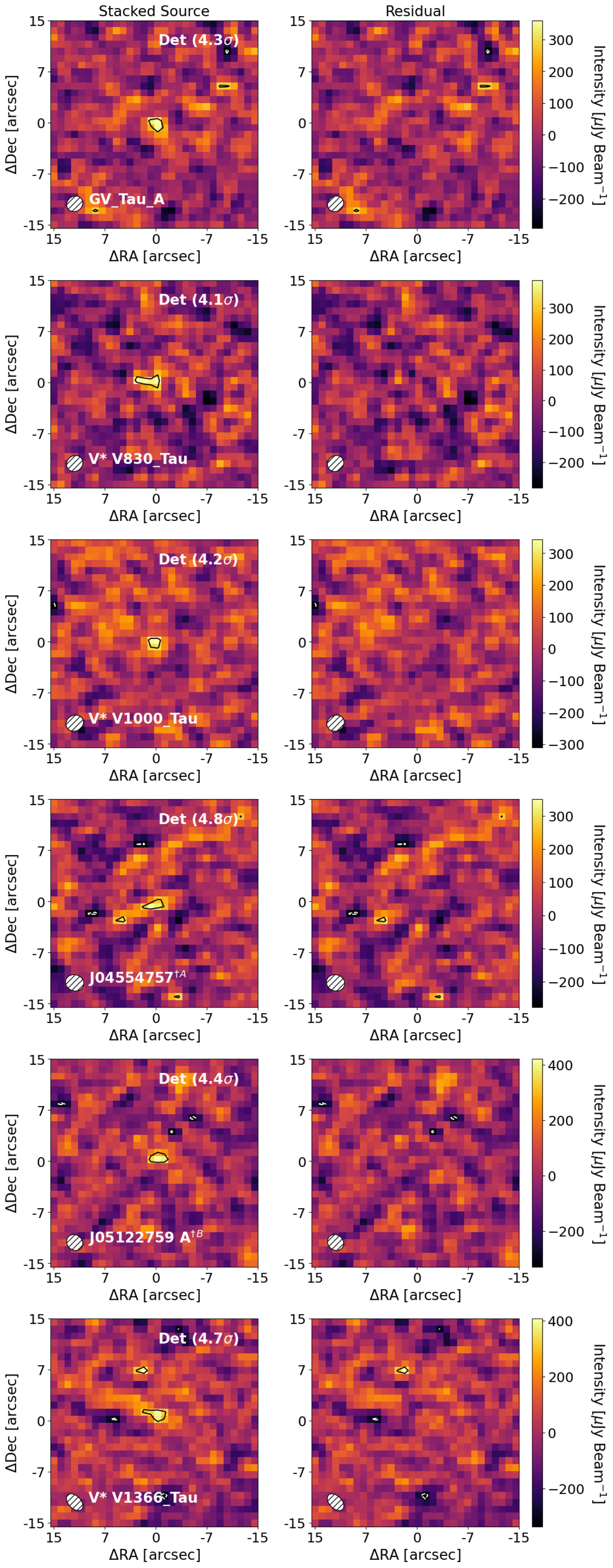}
    \end{minipage}
 \end{figure}
\begin{figure} [t!]
  \caption{Stacked VLASS Quick Look images of detections showing GV Tau A, V* V830 Tau, V* V1000 Tau, J04554757$^\dag$$^A$, J05122759 A$^\dag$$^B$, V* V1366 Tau, as well as their residual maps. VLA beams are indicated in the lower left corners. Contours are at the [-3,3,6,9]$\sigma$ levels.}\label{fig:stacked_dets}
\end{figure}

\section{\textbf{Distance-independent luminosities}}
\label{appA1}
To investigate the luminosity distributions of the VLASS sample, we calculate $F^{\prime}_{\rm{140pc}}$ via two methods, one epoch-averaged, and the other, per-epoch.
In the case of the epoch-averaged value, each of the 512 sources have $F^{\prime}_{\rm{140pc}}$ values derived from their average values in both epochs 1 and 2.
Sources detected in two epochs are simply measured from the mean of the two detected values, whereas for data that either has one or two upper-limit values, the upper-limit values are measured from the mid-point of the upper-limit flux and 0\,$\mu$Jy. 
In deriving these mean values based on one or two upper-limits, our method implicitly assumes that the distribution of fluxes below their upper-limits are symmetric (and uniform). Other assumptions could be made as to the underlying flux distribution of non-detections, however by considering the full range from 0\,$\mu$Jy to the upper-limit fluxes, we consider the conclusions that we derive from $F^{\prime}_{\rm{140pc}}$ to be sound. 
Stacked detections are excluded from this distribution since these represent averages over two epochs.
We note that in our Kaplan-Meier estimators, only sources with non-detections in both epochs are left-censored.
In the case of the per-epoch $F^{\prime}_{\rm{140pc}}$ values, all 512 sample sources explicitly have two unique $F^{\prime}_{\rm{140pc}}$ values that are treated independently (in this case, stacked detections are considered as non-detections).
For these derivations, source detections yield $F^{\prime}_{\rm{140pc}}$ directly from their fluxes, whereas upper-limit values are again derived from the mid-point of the upper-limit flux and 0\,$\mu$Jy.

\section{\textbf{X--Ray data}}
\label{App:xray}
To investigate how the detected sample's radio luminosity densities compare to their X-ray emission, we collect all available published X-ray data for the full sample.
We collate these from the catalogs of \citet{Freund2022}, \citet{Webb2023}, \citet{Rosen2016} and \citet{Gudel2007} and present these in Table~\ref{tab:xraytable}, noting that these X-ray studies conducted the cross-match with known YSOs.
Overall we find 26/33 of the full sample have X--ray detections.
For the \citet{Gudel2007} sample we re-scaled the luminosities from their assumed 140\,pc distance to the new {\it Gaia}-determined parallax distances \citep[provided at the eDR3 epoch in][]{Krolikowski2021}.
In all other samples we applied $L_X = 4 \pi d^2 F_X$, using the published `Flux8', `FX' and `PN8' values for $F_X$ and their {\it Gaia}-determined parallax distances \citep[also provided at the eDR3 epoch in][]{Krolikowski2021}.
We present these alongside the measured radio luminosity densities (calculated from either the mean radio flux of sources detected in both epochs, or from the mean of , i.e., $L_R = 4 \pi d^2 F_R$) for the purpose of assessing the full sample's consistency with the `G\"udel--Benz relationship' \citep{Benz93, Benz94}.
The 26 sources with X-ray data include the 7 sources that we found to be significantly variable, which we exclude from further analysis given that GBR is only valid for quiescent radio emission.

\begin{figure}
    \centering
    \includegraphics[trim={0.1cm 0.1cm 0.0cm 0.1cm},clip,width=0.6\textwidth]{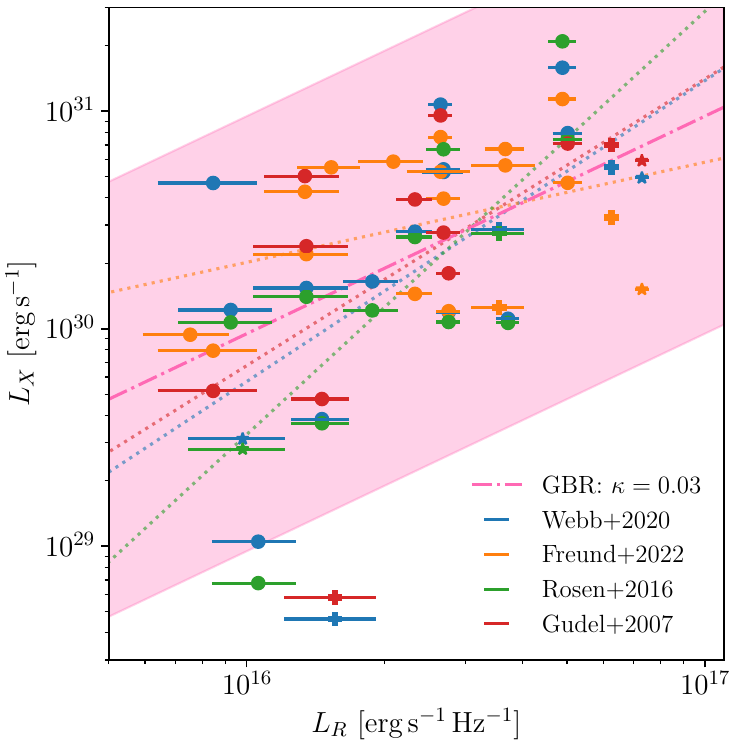}
    \caption{X-ray luminosity versus radio luminosity density for our sample with available X--ray data in the catalogs of \citet{Freund2022}, \citet{Webb2023}, \citet{Rosen2016} and \citet{Gudel2007}, from Table~\ref{tab:xraytable}. Colors indicate each catalog, and markers indicate YSO class (stars: class~0/I/F, pluses: class~II, filled-os: class~III). We overplot the G\"udel-Benz Relationship for $\kappa=0.03$ (in dash-dot hot-pink, plus/minus error, shaded spanning $\pm$ 1 order of magnitude).}
    \label{fig:GBR}
\end{figure}

Fig.~\ref{fig:GBR} presents all available X-Ray data for VLASS-detected sources as a function of their radio luminosity densities.
Given the systematic differences between the X-ray data between the four catalogs, we present these data points independently (with sources having 1--4 separate X--ray measurements).
There is a weak, positive trend between the corresponding X--ray luminosities ($L_X$) and radio luminosity densities ($L_R$).
Further, we calculate $L_R$ values consistent with the G\"udel-Benz Relationship \citep[herein `GBR;'][]{Benz93, Benz94} assuming a $\kappa=0.03$ constant of proportionality, i.e., $L_X {<} \kappa L_R \times10^{15.5{\pm}0.5}$ \citep[the same $\kappa$ value has been found in previous radio YSO surveys, see e.g.,][]{Dzib13, Dzib2015}.
By fitting power-law models to each of the four X-Ray data sets, we also demonstrate these separate data sets are all independently consistent with these G\"udel-Benz Relationship parameters (see dotted lines on Fig.~\ref{fig:GBR}).
We note the presence of one (or two possible) outlier/s beyond the range expected by the GBR (HK~Tau~A and less significantly J04554757+3028077) which are both over-luminous at radio wavelengths.

\begin{table}[h!]
    \centering
    \begin{tabular}{|l|c|c|c|c|c|c|}
    \hline
    ID&Distance& $L_{R(3\,\rm{GHz})}$ & $L_{X1}$ & $L_{X2}$ & $L_{X3}$ & $L_{X4}$ \\
    &[pc]& erg\,s$^{-1}$\,Hz$^{-1}$ & erg\,s$^{-1}$ & erg\,s$^{-1}$ & erg\,s$^{-1}$ & erg\,s$^{-1}$ \\
    \hline
    HD~285281 &136& 1.04E+16 & -- & 4.26E+30 & -- & 5.04E+30 \\
    V1195~Tau~B &176& 2.06E+16 & -- & 5.27E+30 & -- & -- \\
    J0407350+2237396 &125& 8.91E+15 & -- & -- & -- & -- \\
    V773~Tau &120& 6.26E+16 & 5.52E+30 & 3.25E+30 & -- & 6.98E+30 \\
    V410~Tau &129& 2.69E+16 & 5.41E+30 & 3.97E+30 & 6.69E+30 & 2.77E+30 \\
    V*~V1023~Tau &127& 2.33E+16 & 2.80E+30 & 1.45E+30 & 2.64E+30 & 3.93E+30 \\
    V819~Tau &129& 2.45E+16 & 1.19E+30 & 1.20E+30 & 1.07E+30 & 1.80E+30 \\
    HD~283572 &126& 2.65E+16 & 1.07E+31 & 7.61E+30 & -- & 9.57E+30 \\
    T~Tau &144& 7.29E+16 & 4.94E+30 & 1.52E+30 & -- & 5.92E+30 \\
    LkCa~21 &117& 1.46E+16 & 3.84E+29 & -- & 3.68E+29 & 4.75E+29 \\
    V1201~Tau &160& 1.59E+16 & -- & 5.87E+30 & -- & -- \\
    DG~Tau &125& 6.99E+15 & 3.12E+29 & -- & 2.80E+29 & -- \\
    HK~Tau~A &130& 1.22E+16 & 4.63E+28 & -- & -- & 5.81E+28 \\
    CFHT~Tau~12 &175& 1.46E+16 & -- & -- & -- & -- \\
    HD~28867~A &157& 4.89E+16 & 1.59E+31 & 1.14E+31 & 2.10E+31 & -- \\
    XEST~08-003 &161& 2.22E+16 & 5.26E+30 & -- & -- & -- \\
    HP~Tau~G2 &166& 5.02E+16 & 7.94E+30 & 4.69E+30 & 7.43E+30 & 7.10E+30 \\
    J04363248+2421395 &157& 1.22E+16 & -- & -- & -- & -- \\
    HV~Tau~A &137& 3.36E+16 & 1.11E+30 & -- & 1.06E+30 & -- \\
    J04411296+1813194~A &155& 1.20E+16 & -- & -- & -- & -- \\
    LkHa~332~G2 &145& 9.89E+15 & 1.54E+30 & 2.20E+30 & 1.40E+30 & 2.40E+30 \\
    HD~283782 &204& 2.86E+16 & -- & 5.64E+30 & -- & -- \\
    HD~30171 &167& 3.17E+16 & -- & 6.70E+30 & -- & -- \\
    DQ~Tau &195& 3.00E+16 & 2.87E+30 & 1.25E+30 & 2.75E+30 & -- \\
    HD~31305 &143& 1.49E+16 & 1.65E+30 & -- & 1.21E+30 & -- \\
    CIDA~10 &174& 1.58E+16 & -- & -- & -- & -- \\
    LH98~175 &139& 1.20E+16 & -- & 5.52E+30 & -- & -- \\
    LH98 211 &102& 6.35E+15 & -- & -- & -- & -- \\
    \hline
    GV~Tau &142& 9.23E+15 & 1.22E+30 & -- & 1.07E+30 & -- \\
    V*~V830~Tau &130& 8.46E+15 & 4.68E+30 & 7.93E+29 & -- & 5.18E+29 \\
    J04554757+3028050 &146& 1.06E+16 & 1.06E+29 & -- & 6.77E+28 & -- \\
    J05122759+2253492~A &173& 1.47E+16 & -- & -- & -- & -- \\
    V*~V1366~Tau &118& 7.53E+15 & -- & 9.39E+29 & -- & -- \\
    \hline
    \end{tabular}
    \caption{Collated X--ray luminosity and radio luminosity density values (all to 2 S.F.) for the 28 (+5) Taurus detections in our sample for analysis (including the stacked detections in the lower--half of the table). The subscripts 1, 2, 3 and 4 correspond to the public catalogs of \citet{Webb2023}, \citet{Freund2022}, \citet{Rosen2016} and \citet{Gudel2007} respectively. We provide here the \textit{Gaia} parallax distance values used in this work \citep[i.e., from the EDR3 release, as per][]{Krolikowski2021}. }
    \label{tab:xraytable}
\end{table}

\newpage
\nocite{Nguyen2012,Herczeg2014,Kenyon1995,Wichmann1996,Guieu2006,Walter2003,Rebull2010,Luhman2017,Mooley2013,Li1998,White2004}
\bibliography{radiohead}{}
\bibliographystyle{aasjournal}

\end{document}